\documentclass{article}
\usepackage{amsthm,amsmath,amssymb,bm}
\usepackage{mathrsfs}
\usepackage{float} 
\usepackage{graphicx}
\usepackage{footnote}
\usepackage{amsthm,amsmath,amssymb,bm}
\usepackage{mathrsfs}
\usepackage{float} 
\usepackage{graphicx}
\usepackage{footnote}
\usepackage{fdsymbol}
\usepackage{multirow}
\usepackage{cite}
\usepackage[authoryear]{natbib}
\usepackage{algorithm}
\usepackage{algorithmic}
\usepackage{threeparttable}
\usepackage{lscape}

\usepackage{amsthm,amsmath,amssymb,bm}
\usepackage{mathrsfs}
\usepackage{float} 
\usepackage{graphicx}
\usepackage{subfigure}
\usepackage{float}
\usepackage{footnote}
\usepackage{cite}
\usepackage{soul}
\usepackage{authblk}
\usepackage{color,xcolor}
\usepackage{amsmath,amsfonts,amssymb,amsthm,booktabs,color,epsfig,graphicx,hyperref,url}
\usepackage{mathrsfs}
\usepackage{float} 
\usepackage{graphicx}
\usepackage{footnote}
\usepackage{fdsymbol}
\usepackage{cite}
\usepackage{multirow}
\usepackage{subfigure}
\usepackage{algorithm}
\usepackage{algorithmic}
\usepackage{soul}
\usepackage{color,xcolor}
\soulregister{\citet}7
\soulregister{\romannumeral}7
\sethlcolor{yellow}

\begin{document}
	
	\title{A New Covariate Selection Strategy for High Dimensional Data in Causal Effect Estimation with Multivariate Treatments}
	\author[1]{Juan Chen}
	\author[1]{Yingchun Zhou \thanks{Corresponding author:  yczhou@stat.ecnu.edu.cn}}
	\affil[1]{Key Laboratory of Advanced Theory and Application in Statistics
		and Data Science-MOE, School of Statistics, East China Normal University.}
	\date{}
	
	\maketitle
	\begin{abstract}
		Selection of covariates is crucial in the estimation of average treatment effects given observational data with high or even ultra-high dimensional pretreatment variables. Existing methods for this problem typically assume sparse linear models for both outcome and univariate treatment, and cannot handle situations with ultra-high dimensional covariates. In this paper, we propose a new covariate selection strategy called double screening prior adaptive lasso (DSPAL) to select confounders and predictors of the outcome for multivariate treatments, which combines the adaptive lasso method with the marginal conditional (in)dependence prior information to select target covariates, in order to eliminate confounding bias and improve statistical efficiency. The distinctive features of our proposal are that it can be applied to high-dimensional or even ultra-high dimensional covariates for multivariate treatments, and can deal with the cases of both parametric and nonparametric outcome models, which makes it more robust compared to other methods. Our theoretical analyses show that the proposed procedure enjoys the sure screening property, the ranking consistency property and the variable selection consistency. Through a simulation study, we demonstrate that the proposed approach selects all confounders and predictors consistently and estimates the multivariate treatment effects with smaller bias and mean squared error compared to several alternatives under various scenarios. In real data analysis, the method is applied to estimate the causal effect of a three-dimensional continuous environmental treatment on cholesterol level and enlightening results are obtained.
		
	\end{abstract}
	\textbf{Keywords}: 	adaptive lasso; causal effect; covariate selection; multivariate continuous treatments
	
	\newpage
	\section{Introduction\label{sec:1}}
	
	The main challenge of estimating causal effect from observational data is the existence of confounders that are associated with both treatment and outcome, which may lead to biased causal estimates. To remove such bias, the propensity score (PS), defined as the conditional probability of assignment to a particular treatment given covariates, is commonly used to control the confounding effect (\citet{rosenbaum1983central}; \citet{rosenbaum1984reducing}; \citet{rosenbaum1985constructing}; \citet{robins2000marginal}; \citet{hirano2004propensity}). An important assumption of the propensity score based methods is that all confounders are measured and included in the propensity score model. Indeed, researches have shown that inclusion of unnessary covariates and exclusion of important confounders can lead to biased causal effect estimates and efficiency loss. Hence it is important to determine which covariates should be included into the PS model and the outcome model so as to estimate causal effect consistently and efficiently. A typical way of covariate selection is driven by expert knowledge, which becomes difficult when the covariates are high dimensional or even ultra-high dimensional.  
	\noindent
	
	In response to this challenge, there has been a lot of studies on developing data-driven procedures for covariate selection in causal inference. These methods generally divide covariates into four disjoint subsets under the framework of causal directed acyclic graph (DAG) (\citet{pearl2009causality}):
	\begin{enumerate}
		\item Instrumental variables ($X_I$), which are related to treatment but not to outcome, unless through treatment;
		\item Counfounders ($X_C$), which are related to both treatment and outcome; 
		\item  Outcome predictors ($X_P$), which predict outcome only;
		\item Spurious variables ($X_S$), which are not related to treatment or outcome.
	\end{enumerate}
	Figure 1 provides the simplest causal diagram associated with these definitions.
	\begin{figure}[h]
		\centering
		\includegraphics[scale=0.5]{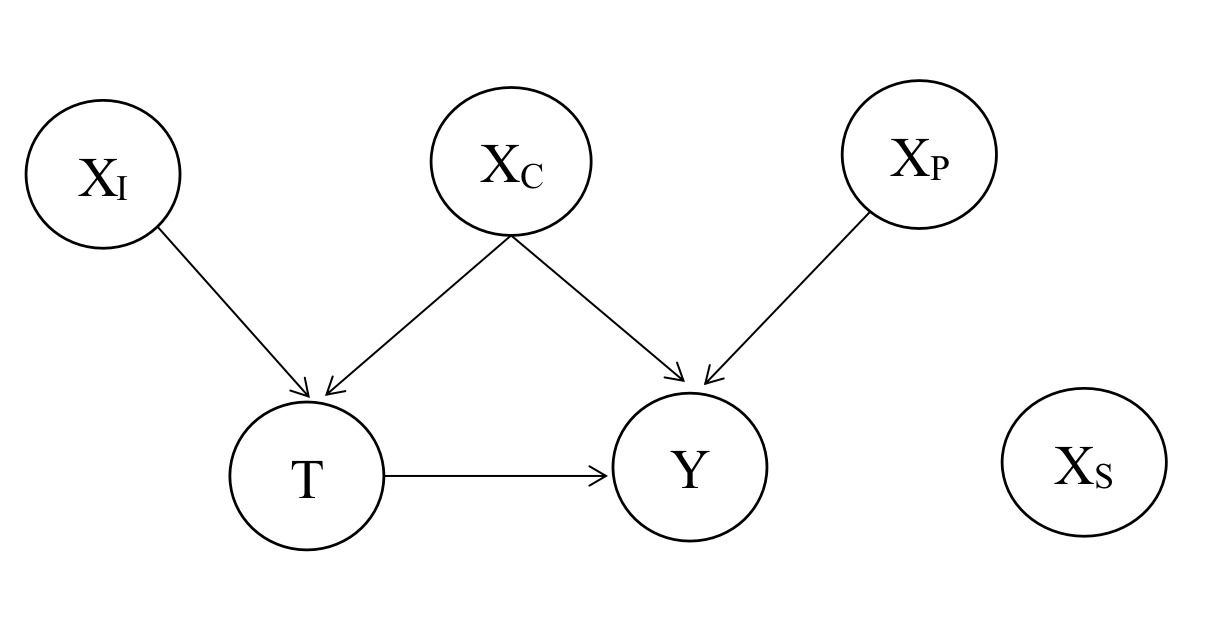}
		\caption{A causal directed acyclic graph demonstrating four types of covariates}
	\end{figure}
	Researches have shown that inclusion of instrumental variables in the covariates can lead to variance inflation (\citet{de2011covariate} and \citet{patrick2011implicationsf}), and inclusion of outcome predictors may lead to efficiency gains (\citet{brookhart2006variable}). This implies that an efficient covariate selection method should achieve the optimal covariate set, which includes both confounders and predictors of the outcome and excludes instrumental variables, i.e., the target covariate set is $X_C \cup X_P$. 
	\noindent
	
	Motivated by these results, various procedures have been developed. For univariate binary treatment, \citet{ertefaie2018variable} proposed a variable selection method using a penalized objective function based on a linear outcome model and a logistic propensity score model. \citet{wang2020debiased} proposed a debiased inverse propensity score weighting (DIPW) scheme for average treatment effect estimation when the propensity score follows a sparse logistic regression model. Besides, there are some literature using Bayesian methods to select covariates such as the Bayesian confounding adjustment method introduced by \citet{antonelli2019high}. For univariate multivalued treatment, \citet{farrell2015robust} developed a group lasso method based on a linear outcome model and a multinomial logistic propensity score model. For univariate continuous treatment, \citet{antonelli2020causal} proposed a Bayesian framework for estimating causal effect curve of a continuous treatment when the covariates are high dimensional. For multivariate continuous treatments, under a sparse linear outcome model, \citet{wilson2018model} developed a Bayesian model-averaging method to estimate the effect of a multivariate exposure on an outcome when the covariates are high dimensional. However, the validity of this method relies on the correct specification of the linear outcome regression model. None of these methods are suitable for situations with ultra-high dimensional covariates and most of these methods cannot be applied to multivariate continuous treatments. 
	\noindent
	
	To overcome these difficulties, this paper proposes a Double Screening Prior Adaptive Lasso (DSPAL) method that first screens out all the confounders and instrumental variables by an independence screening procedure, then a covariate set that includes all confounders and outcome predictors is selected by a conditional independence screening procedure based on the covariates selected in the first step. Next is a major step that combines the adaptive lasso method with prior weight, which is constructed from the conditional screening statistic in the second step, to further select the target covariates. Finally, the causal effect function of multivariate continuous treatments is estimated based on the covariates selected. The DSPAL method is well suitable for causal effect estimation for multivariate continuous treatments with high-dimensional or even ultra-high dimensional covariates. Besides, it can be used in the case of both parametric and nonparametric outcome model, which makes it more robust compared to other methods. Theoretical properties and simulation results indicate that DSPAL selects all confounders and predictors consistently and outperforms other methods under various scenarios.
	\noindent
	
	The remainder of this article is organized as follows. Section 2 introduces the notation and assumptions. Section 3 introduces the proposed DSPAL method. Section 4 establishes the theoretical properties of the DSPAL method. Section 5 performs numerical simulation under various scenarios and compare the performance of different methods. Section 6 applies the proposed method to a real data analysis and the conclusion of this article is drawn in Section 7.
	
	\section{Notation and Assumptions \label{sec:2}}
	Suppose the treatment for subject $i$ is $\mathbf{T}_i = (T_{i1},\dots, T_{iq})^{'}$, whose support is $\mathcal{T} \subset \mathcal{R}^q$. $\mathbf{X}_i = (X_{i1},\dots, X_{ip})^{'} \in \mathcal{R}^p$ denotes the observed covariates, where $p,q$ denote the dimensions of the covariates and treatments, respectively. Suppose that for each subject, there exists a potential outcome $Y_i(\mathbf{t})$ for all $\mathbf{t} \in \mathcal{T}$. The observed outcome is defined as $Y_i = Y_i(\mathbf{t})$ if $\mathbf{T}_i = \mathbf{t}$. Assume that a sample of observations $\lbrace Y_i, \mathbf{T}_i, \mathbf{X}_i \rbrace$ for $i \in \lbrace 1,....,n \rbrace$ is independently drawn from a joint distribution $f(Y,\mathbf{T}, \mathbf{X})$. For notational convenience, the treatments and covariates are assumed to be standardized.
	
	\noindent 
	
	To perform causal inference with observational data, three standard assumptions are made (\citet{hirano2004propensity}; \citet{imai2004causal}):
	\vspace{12pt}
	
	\noindent
	\textbf{Assumption 1 (Ignoribility)}:\\
	$\mathbf{T}_i \perp Y_i(\mathbf{t}) \mid \mathbf{X}_i$, meaning that the treatment assignment is independent of the potential outcomes given covariates.
	
	\vspace{12pt}
	
	\noindent
	\textbf{Assumption 2 (Positivity)}:\\
	$f_{\mathbf{T} \mid \mathbf{X}}(\mathbf{T}_i = \mathbf{t} \mid \mathbf{X}_i ) > 0$ for all $\mathbf{t} \in \mathcal{T}$, where the conditional density $f(\mathbf{T}_i \mid \mathbf{X}_i )$ is called the generalized propensity score (\citet{imbens2000role}).
	\vspace{12pt}
	
	\noindent
	\textbf{Assumption 3 (SUTVA)}:\\
	There is no interference among units, which means that each individual's outcome only depends on their own level of treatment intensity.
	
	\section{Double Screening Prior Adaptive Lasso for Multivariate Continuous Treatments \label{sec:3}}
	
	In this section, the Double Screening Prior Adaptive Lasso (DSPAL) method is introduced, which estimates the causal effect function of multivariate continuous treatments when the covariates are high dimensional or even ultra-high dimensional. Firstly, an independence screening procedure is carried out to screen out all the confounders and instrumental variables. This procedure is based on the canonical correlation statistic that can be applied to feature screening for multivariate continuous treatments. Then a covariate set that includes all confounders and outcome predictors is selected by a conditional independence screening procedure based on the generalised covariance measure statistic. Next is a major step that combines the adaptive lasso method with the prior weight, which is constructed from the generalised covariance measure statistic in the previous step, to exclude the spurious variables and instrumental variables that might remain. Finally, the causal effect function of multivariate continuous treatments is estimated based on the covariates selected.
	\noindent
	
	Specifically,  it is assumed that the treatment assignment model is a multiple multivariate linear model. Let  $\mathbf{\epsilon}_i = (\epsilon_{i1},\dots,\epsilon_{iq})^{'}$ denote the errors, then the multiple multivariate linear treatment assignment model is 
	\begin{equation}
		\mathbf{T}_i = \mathbf{B}^{'}\mathbf{X}_i+\mathbf{\epsilon}_i \ \mbox{for} \ i \in \lbrace 1,....,n \rbrace,
	\end{equation}
	where $\textbf{B}$ denotes a $p \times q$ coefficient matrix and $\mathbf{\epsilon}_i \mathop{\sim} \limits^{i.i.d} N_q(\mathbf{0},\mathbf{\Sigma})$.
	\noindent
	
	The model can also be expressed in matrix form, which is
	\begin{equation}
		\mathbf{T} = \mathbf{X}\mathbf{B}+\mathbf{E},
	\end{equation}
	where $\textbf{X}$ denotes the $n \times p$ predictor matrix, whose $i$th row is $\textbf{X}_i^{'}$; $\textbf{T}$ denotes the $n \times q$ treatment matrix, whose $i$th row is $\textbf{T}_i^{'}$ and $\textbf{E}$ denotes the $n \times q$ error matrix, whose $i$th row is $\mathbf{\epsilon}^{'}_i$.
	\noindent
	
	This regression model has $pq$ parameters to be estimated, which becomes challenging when the covariates are high dimensional or ultra-high dimensional. To meet this challenge, we first develop a double screening procedure to reduce the dimension of covariates, and then combine the adaptive lasso method with the marginal conditional (in)dependence prior information to select all confounders and predictors.

	\subsection{Independence screening}
	\noindent
	
	When the candidate covariate set is high dimensional or ultra-high dimensional, a common strategy is to use the sure independence screening procedure based on marginal correlations (\citet{fan2008sure}) or conditional correlations (\citet{barut2016conditional}). From Figure 1, one can see that if the DAG is faithful (\citet{pearl2009causality}), then the following (in)dependence can be obtained:
	\begin{equation}
		X_C \nVbar T, \ X_I \nVbar T,\ X_P \Vbar T, \ X_S \Vbar T.
	\end{equation}
	Based on the (in)dependence, one can select a covariate set containing $X_C$. In this paper, we use multi-treatment canonial correlation, which is proposed by \citet{di2021sure}, to perform independence screening based on (3) and the notion of multi-treatment canonial correlation is first introduced.
	
	
	Define the multi-treatment canonial correlation $r_k^c$  (\citet{di2021sure}) between $X_k$ and $\textbf{T} = (T_1,\dots,T_q)$ as
	\begin{equation}
		r_k^c = max_\textbf{b} \frac{\Sigma_{X_k\textbf{T}}\textbf{b}}{\sqrt{\sigma_{k}^2}\sqrt{\textbf{b}^{'}\Sigma_\textbf{T}\textbf{b}}},
	\end{equation}
	where $\sigma_{k}^2,\Sigma_{X_k\textbf{T}}, \Sigma_\textbf{T}$ are the submatrices of $\Sigma = \begin{bmatrix}
		\sigma_{k}^2	 & \Sigma_{X_k\textbf{T}} \\
		\Sigma_{\textbf{T}X_k}&  \Sigma_\textbf{T}
	\end{bmatrix}$, which is the covariance matrix of $(X_k,(T_1,\dots,T_q))$. 
	One can show that 
	\begin{equation}
		(r_k^c)^2 = (\mathbf{r}_k)^{'}(\mathbf{\Psi_\textbf{T}})^{-1} (\mathbf{r}_k),
	\end{equation}
	where $\mathbf{r}_k = (r_{k1},\dots,r_{kq})^{'}$ are the Pearson correlations between $X_k$ and $T_j^{'}s$, and $\mathbf{\Psi_\textbf{T}} = (\psi_{kl})_{q\times q}$ is the correlation matrix of $(T_1,\dots,T_q)$. Therefore,  
	\begin{equation}
		\hat{r}_k^c =  (\hat{\mathbf{r}}_k)^{'}(\hat{\mathbf{\Psi}}_\textbf{T})^{-1} (\hat{\mathbf{r}}_k)
	\end{equation}
	is used as the canonical correlation screening statistic. In practice, one can pick the top $D$ many variables according to the $D$ $\hat{r}_k^c $ values. The selected covariates set is denoted as $\mathcal {M}_C$.
	\subsection{Conditional independence screening}
	\noindent
	
	From Figure 1, one can see that if the DAG is faithful (\citet{pearl2009causality}), then the following conditional (in)dependence can be obtained:
	\begin{gather}
		X_I \Vbar Y \mid (T \cup X_C),
		\ X_P \nVbar Y \mid (T \cup X_C), \\  \notag
		\ X_S \Vbar Y \mid (T \cup X_C), 
		\ X_j \nVbar Y \mid (T \cup X_{C(-j)}) \ \mbox{for} \ j \in C.
	\end{gather}
	The reason for the extra control of $X_C$ is that $T$ is the collider in the path $X_I \to T \gets X_C \to Y$, and controlling only $T$ will lead to collider bias since $X_I$ can influence Y through this path, hence we need to control $T$ and $X_C$ jointly. It seems that conditional independence screening based on condition (7) can be used to select $X_C \cup X_P$. In practice, it becomes difficult  since $X_C$ is unknown. However, the conditional set $X_C$ can be repalced by any set that contains $X_C$ if Figure 1 is faithful. Therefore, $X_C$ in (7) is replaced by $\mathcal{M}_C$ selected in the independence screening step, which leads to the following contidional independence:
	
	\begin{equation*}
		X_C \nVbar Y \mid (T \cup \mathcal {M}_C), \ X_I \Vbar Y \mid (T \cup \mathcal {M}_C), 
	\end{equation*}
	\begin{equation}
		\ X_P \nVbar Y \mid (T \cup \mathcal {M}_C),\ X_S \Vbar Y \mid (T \cup \mathcal {M}_C).
	\end{equation}
	
	\noindent
	
	To perform conditional independence screening based on condition (8), we first introduce the notion of the generalised covariance measure (GCM) statistic, which is described in detail in \citet{shah2020hardness}. The generalised covariance measure statistic between $X$ and $Y$ given $Z$ is defined as the normalized covariance between the residuals from the regression models of $Y$ and $X$ on $Z$, respectively. Here considers the univariate $X$ and $Y$ given multivariate $\mathbf{Z}$.
	
	Given a distribution $P$ for $(X,Y,\mathbf{Z})$, assume
	\begin{equation*}
		X=f(\mathbf{Z})+\zeta_x, \ Y=g(\mathbf{Z})+\zeta_y,
	\end{equation*} 
	where $f(\mathbf{z}) = \mathbb{E}(X \mid \mathbf{Z}=\mathbf{z})$ and  $g(\mathbf{z}) = \mathbb{E}(Y \mid \mathbf{Z}=\mathbf{z})$. Let $\hat{f}(\mathbf{z})$ and  $\hat{g}(\mathbf{z})$ be the estimates of $f(\mathbf{z})$ and $g(\mathbf{z})$, then the product between residuals from the regression models is:
	\begin{equation*}
		R_i = (x_i-\hat{f}(\mathbf{z_i}))(y_i-\hat{g}(\mathbf{z_i})),\ \forall i=1,\dots,n.
	\end{equation*}
	The generalised covariance measure statistic $GCM$ is defined as:
	\begin{equation}
		\hat{GCM} = \frac{\frac{1}{n}\sum_{i=1}^{n}R_i}{(\frac{1}{n}\sum_{i=1}^{n}R_i^2-(\frac{1}{n}\sum_{i=1}^{n}R_i)^2)^{1/2}}
	\end{equation}
	
	Since large values of $\mid \hat{GCM} \mid$ suggests rejecting the hypothesis $H_0: X \Vbar Y \mid \mathbf{Z}$, one can use $\mid \hat{GCM} \mid$ as a conditional independence screening statistic to select the target covariates. Therefore, given a pre-specified threshold $K$, consider 
	\begin{equation*}
		\mathcal {M}_{C \cup P} = \lbrace k: \hat{GCM}_{k} \ \text{is among the top} \ K \ \text{largest of all} \rbrace
	\end{equation*}
	as an estimator of the active set $	\mathcal {M}^*_{C \cup P} $, where
	\begin{equation*}
		\mathcal {M}^*_{C \cup P} = \lbrace k: X_k \ \text{conditionally depends on}\ Y  \ \text{given}\  \mathbf{Z} \rbrace.
	\end{equation*}
	In this paper, as suggested by \citet{fan2008sure}, $K= \lfloor n/\text{log}(n) \rfloor$. Besides, $\hat{GCM}$ are obtained via kernel ridge regression, which is proposed by \citet{shah2020hardness}.

	\subsection{Prior adaptive Lasso procedure}
	\noindent
	
	Since there may be some instrumental variables ($X_I$) and spurious variables ($X_S$) remaining in the selected set $\mathcal {M}_{C \cup P}$ after the second step, a third variable selection step is proposed to further exclude these variables. In the usual adaptive lasso method, the weights for $X_S$ and $X_P$ are large and those for $X_C$ and $X_I$ are small, therefore in order to select $X_C \cup X_P$, the idea is to impose a heavier weight on $X_I$ than $X_P$ in the penalty terms using the multiple multivariate linear treatment assignment model introduced in subsection 2.2. The challenge is to specify appropriate weights to select covariates consistently. A new method that combines the adaptive lasso method with the marginal conditional (in)dependence prior information is proposed to select the target covariates, which can select $X_C \cup X_P$ consistently for multivariate continuous treatments.
	
	Specifically, the negative log-likelihood function of the treatment assignment model is 
	\begin{equation}
		L(\mathbf{B},\mathbf{\Sigma})= \mbox{tr}\lbrace \frac{1}{n}(\mathbf{T}-\mathbf{X}\mathbf{B})^{'}(\mathbf{T}-\mathbf{X}\mathbf{B})\mathbf{\Sigma}^{-1}  \rbrace-\mbox{ln}\mid \mathbf{\Sigma}^{-1} \mid.
	\end{equation}
	Let $\mathbf{\Omega} = \mathbf{\Sigma}^{-1}$, then 
	\begin{equation}
		L(\mathbf{B},\mathbf{\Omega})= \mbox{tr}\lbrace \frac{1}{n}(\mathbf{T}-\mathbf{X}\mathbf{B})^{'}(\mathbf{T}-\mathbf{X}\mathbf{B})\mathbf{\Omega} \rbrace-\mbox{ln}\mid \mathbf{\Omega} \mid.
	\end{equation}
	One penalty is added to the negative log-likelihood function $L(\mathbf{B},\mathbf{\Omega})$ to construct the sparse estimator of $\mathbf{B}$:
	\begin{gather}
		(\hat{\mathbf{B}}, \hat{\mathbf{\Omega}})  = \text{argmin}_{\mathbf{\Omega},\mathbf{B}}  \lbrace \text{tr}\lbrace \frac{1}{n}(\mathbf{T}-\mathbf{X}\mathbf{B})^{'}(\mathbf{T}-\mathbf{X}\mathbf{B})\mathbf{\Omega} \rbrace-\text{ln}\mid \mathbf{\Omega}  \mid \notag \\  
		+ \lambda \sum_{i=1}^{p}\sum_{j=1}^{q}\hat{w}_{ij}\mid B_{ij} \mid \rbrace,
	\end{gather}
	\noindent
	where $\hat{w}_{ij} = ( \frac{\mid \hat{GCM}_{i} \mid}{\text{max}_i \mid \hat{GCM}_{i} \mid} )^{-\gamma}, \  \forall j, \gamma>0$ and 
	$\lambda$ is a tuning parameter. 
	This method is called as prior adaptive lasso (PAL). Unlike usual adaptive lasso, the PAL imposes a heavier penalty weight on the covariates that are less associated with outcome implied by the result of Step 2. Therefore, it is more likely to select $X_C$ and $X_P$ than $X_I$ and $X_S$. 

	Note that the optimization problem in equation (12) is not convex. However, it would be a convex problem if either $\mathbf{B}$ or $\mathbf{\Omega}$ is fixed, hence an iteration algorithm is proposed.
	First estimate $\mathbf{\Omega}$ with fixed $\mathbf{B}$ through optimize the following function:
	\begin{equation}
		\hat{\mathbf{\Omega}}(\mathbf{B}) = \text{argmin}_\mathbf{\Omega} \lbrace \text{tr}(\frac{1}{n}(\mathbf{T}-\mathbf{X}\mathbf{B})^{'}(\mathbf{T}-\mathbf{X}\mathbf{B})\mathbf{\Omega})- \text{ln}\mid \mathbf{\Omega} \mid  \rbrace.
	\end{equation}
	This covariance estimation problem can be solved by the maximum likelihood method.
	
	Then the estimate of $\mathbf{B}$ based on the updated estimate of $\mathbf{\Omega}$ can be obtained, which yields the following optimization problem:
	\begin{equation}
		\hat{\mathbf{B}}(\mathbf{\Omega}) = \text{argmin}_{\mathbf{B}} \lbrace \text{tr}[\frac{1}{n}(\mathbf{T}-\mathbf{X}\mathbf{B})^{'}(\mathbf{T}-\mathbf{X}\mathbf{B})\mathbf{\Omega}]+\lambda\sum_{i=1}^{p}\sum_{j=1}^{q}\hat{w}_{ij}\mid B_{ij} \mid \rbrace
	\end{equation}
	Since the penalized log-likelihood function in equation (14) is quadratic, denoted as $g(\mathbf{B}) = \text{tr}[\frac{1}{n}(\mathbf{T}-\mathbf{X}\mathbf{B})^{'}(\mathbf{T}-\mathbf{X}\mathbf{B})\mathbf{\Omega}]+\lambda \sum_{i=1}^{p}\sum_{j=1}^{q}\hat{w}_{ij}\mid B_{ij} \mid$, a direct coordinate descent algorithm can be used to estimate $\mathbf{B}$. 
	
	Specifically, the derivative of $g(\mathbf{B})$ respect to $B_{ij}$ is 
	\begin{equation}
		\frac{\partial g(\mathbf{B})}{\partial B_{ij}} = \frac{2}{n}\mathbf{e}_j^{'}(\mathbf{X}^{'}\mathbf{X}\mathbf{B}\mathbf{\Omega})\mathbf{e}_i + \lambda \hat{w}_{ij} sgn(B_{ij})- \frac{2}{n}\mathbf{e}_j^{'}(\mathbf{X}^{'}\mathbf{T}\mathbf{\Omega})\mathbf{e}_i,
	\end{equation}
	where $\mathbf{e}_j$ and $\mathbf{e}_i$ are the corresponding base vectors with $p$ and $q$ dimensions, the function $sgn(\cdot)$ is defined as
	\[	sgn(x)= \begin{cases}
		1,	\hspace*{0.5cm} & x >0; \\
		0, 	\hspace*{0.5cm}& x=0;\\
		-1,	\hspace*{0.5cm}& x<0.
	\end{cases} \]
	
	Setting equation (15) to zero, the updated formula for $B_{ij}$ is 
	\begin{equation}
		\hat{B}_{ij} = sgn(h_{ij}) \frac{n(\mid h_{ij} \mid-\lambda\hat{w}_{ij})_+}{2(\mathbf{e}_j^{'}\mathbf{X}^{'}\mathbf{X}\mathbf{e}_j)(\mathbf{e}_i^{'}\mathbf{\Omega}\mathbf{e}_i)},
	\end{equation}
	where $h_{ij} = \frac{2}{n}\lbrace \mathbf{e}_j^{'}(\mathbf{X}^{'}\mathbf{T}\mathbf{\Omega})\mathbf{e}_i + (\mathbf{e}_j^{'}\mathbf{X}^{'}\mathbf{X}\mathbf{e}_j)(\mathbf{e}_i^{'}\mathbf{\Omega}\mathbf{e}_i)\tilde{B}_{ij}-
	\mathbf{e}_j^{'}(\mathbf{X}^{'}\mathbf{X}\tilde{\mathbf{B}}\mathbf{\Omega})\mathbf{e}_i \rbrace$ and $\tilde{\mathbf{B}}$, $\tilde{B}_{ij}$ are the estimates in the previous step of the iteration. The proof of equation (16) can be found in the supplementary materials.
	
	Combining the two steps together, the following iteration algorithm is developed to solve the optimization problem in equation (12).
	
	\begin{algorithm}[htb]
		\caption{The iteration algorithm for solving $\mathbf{B}$ and $\mathbf{\Omega}$: }
		\label{alg:Framwork}
		\begin{algorithmic}[1] 
			\STATE For fixed $\lambda$, start with $\hat{B}_0 = \mathbf{0} $ and $\hat{\mathbf{\Sigma}}_0 = \frac{1}{n}\mathbf{T}^{'}\mathbf{T}$; \\
			\STATE Compute $\hat{\mathbf{\Omega}}^{(k+1)} = \hat{\mathbf{\Omega}}(\hat{\mathbf{B}}^{(k)})$ by solving equation (3.13) using maximum likelihood method;\\
			\STATE  Compute $\hat{\mathbf{B}}^{(k+1)} = \hat{\mathbf{B}}(\hat{\mathbf{\Omega}}^{(k+1)})$ by solving equation (3.14) using updated formula (3.16);\\
			\STATE  If $\sum_{i=1}^{p}\sum_{j=1}^{q} \mid \hat{B}_{ij}^{(k+1)}-\hat{B}_{ij}^{(k)} \mid < \epsilon$, then stop; otherwise turn to Step 2;\\
			\STATE  Output the estimates $\hat{\mathbf{B}}$ and $\hat{\mathbf{\Omega}}$.
		\end{algorithmic}
	\end{algorithm}

	\noindent
	%
	
	For the DSPAL method, the tuning parameters $\lambda$ is selected according to Bayesian information criteria (BIC), which is defined as 
	\begin{equation*}
		\mbox{BIC}(\hat{\mathbf{\Omega}},\hat{\mathbf{B}}) = -n\mbox{log}(\mid \hat{\mathbf{\Omega}} \mid)+n \mbox{tr}\lbrace\frac{1}{n}(\mathbf{T}-\mathbf{X}\hat{\mathbf{B}})^{'}(\mathbf{T}-\mathbf{X}\hat{\mathbf{B}}) \hat{\mathbf{\Omega}}\rbrace+\text{ln}(n)(k+q+s/2),
	\end{equation*}
	where $q$ is the dimension of $\mathbf{T}$, $k$ is the number of non-zero elements of $\hat{\mathbf{B}}$, $s$ is the number of non-zero off-diagonal elements of $\hat{\mathbf{\Omega}}$. The BIC has been shown to perform well for selecting the tuning parameter for conditional Gaussian graphical model (cGGM) (\citet{yin2011sparse}).
	
	\subsection{Causal Effect Estimation for Multivariate Continuous Treatments}
	\noindent
	
	Based on the covariates selected by the DSPAL method, the entropy balancing for multivariate treatments (EBMT) method (\citet{chen2022causal}) is used to obtain the stabilized weight to balance these covariates and then parametric or nonpapametric method is used to estimate the causal effect function based on the stabilized weight. 
	\noindent
	
	The stabilized weight is defined as
	\begin{equation}
		w_i = \frac{f(\mathbf{T}_i)}{f(\mathbf{T}_i \mid \mathbf{X}_i)},
	\end{equation}
	where $f(\mathbf{T}_i)$ and $f(\mathbf{T}_i \mid \mathbf{X}_i)$ are the density density function of $\mathbf{T}_i$ and $\mathbf{T}_i \mid \mathbf{X}_i$, respectively. one can obtain the estimate of $w_i$ by solving the following optimization problem: 
	\begin{gather}
		\text{min}_w \sum_{i=1}^{n}w_i\text{ln}(\frac{w_i}{v_i})  \notag                                   
	\end{gather}
	s.t.
	\begin{gather}
		\sum_{i=1}^{n}w_ig(\mathbf{T}_i,  \mathbf{X}_i) = \mathbf{0}, \	\sum_{i=1}^{n}w_i = 1,  \ w_i > 0  \ \forall i = 1,..., n.  
	\end{gather}
	where $g(\mathbf{T}_i,  \mathbf{X}_i)  = [\text{vec}((\textbf{T}_i\textbf{X}_i^{'}))^{'},\textbf{T}_i,  \textbf{X}_i]^{'}$, $v_i$ is the base weight and equals to $1/n$ in this paper. Using a method that is similar to the standard Lagrange multiplier technique to solve this optimization problem, one may obtain the weight in terms of the Lagrange multipliers $\gamma$ as 
	\begin{equation}
		w_i = \frac{v_i\text{exp}(-\gamma^{'}g(\mathbf{T}_i,  \mathbf{X}_i))}{\sum_{i=1}^{n}v_i\text{exp}(-\gamma^{'}g(\mathbf{T}_i,  \mathbf{X}_i))}.
	\end{equation}
	where $\gamma$ is a solution of the new dual objective function:
	\begin{equation}
		\text{min}_\gamma \ \text{ln}(\sum_{i=1}^{n}v_i\text{exp}(-\gamma^{'}g(\mathbf{T}_i,  \mathbf{X}_i))).
	\end{equation}
	This new dual objective function can be optimized by using an efficient convex optimization algorithm.
	\noindent
	
	Based on the estimated weight, one can estimate the causal effect function using parametric or nonparametric methods (\citet{chen2022causal}). For parametric causal effect function $\mathbb{E}(Y(\mathbf{t}))= s(\mathbf{t};\theta)$, there is a unique solution  $\theta^* \in \mathcal{R}^p$ defined as
	\begin{equation}
		\theta^* = \text{agrmin}_\theta \int_{\mathcal{T}} \mathbb{E}[Y(\mathbf{t})-s(\mathbf{t};\theta)]^2f_\mathbf{T}(\mathbf{t})d\mathbf{t}.
	\end{equation}
	Under Assumption 1, the true value $\theta^*$ is also a solution for the weighted optimization problem:
	\begin{equation}
		\theta^* = \text{argmin}_\theta  \mathbb{E}[w(Y-s(\mathbf{T};\theta))^2]. 
	\end{equation}
	For nonparametric causal effect function $\mathbb{E}(Y(\mathbf{t}))= s(\mathbf{t})$, one can show that $\mathbb{E}(wY \mid \mathbf{T}=\mathbf{t}) = \mathbb{E}(Y(\mathbf{t}))$, and can use $B$-splines to approximate the causal effect function. 
	
	\section{Large Sample Properties\label{sec:4}}
	
	This section establishes the theoretical properties of the proposed estimators in Section 3. For the double screening stage (Steps 1-2), since the theoretical properties of the independence screening (Step 1) have been shown under some assumptions (Di He et al., 2021, Theorem 1), here will show the theoretical properties of the conditional independence screening (Step 2), which guarantee that the selected set $\mathcal{M}_{C \cup P}$ includes all the confounders and predictors with probability approaching to 1 as $n$ goes to infinity. The following two assumptions are made:
	\noindent
	
	(A1) There exist two positive constants $c >0$ and $0 < \eta \leq 1/2$ such that $\text{min}_{k \in \mathcal{M}^*_{C \cup P}} GCM_k \geq 2cn^{-\eta}$.
	\noindent
	
	(A2) $\text{min}_{k \in \mathcal{M}^*_{C \cup P}} GCM_k - \text{max}_{k \in (\mathcal{M}^*_{C \cup P})^c} GCM_k \geq 0$.
	\noindent
	
	(A3) $(\frac{1}{n}\sum_{i=1}^{n}\lbrace f(z_i) - \hat{f}(z_i) \rbrace^2)(\frac{1}{n}\sum_{i=1}^{n}\lbrace g(z_i) - \hat{g}(z_i) \rbrace^2)=o_p(n^{-1})$ and $0<E(\zeta_x^2\zeta_y^2)<\infty$.
	\noindent
	
	(A4) For fixed $\mathbf{\Omega}$, there exists  function $R(\mathbf{x}), x\in \mathcal{R}^K$ such that for $\mathbf{B}$ in the neighborhood of $\mathbf{B}^*$  satisfying 
	\begin{equation*}
		\mid \phi^{''}(\mathbf{x}, \mathbf{\Omega}, \mathbf{B}) \mid \leq R(\mathbf{x}), 
	\end{equation*}
	where $\phi(\mathbf{x}, \mathbf{\Omega}, \mathbf{B}) = -\text{log} \mid \mathbf{\Omega} \mid +\text{tr} \lbrace (\mathbf{\Omega} (\mathbf{T}-\mathbf{x}\mathbf{B})^{'}(\mathbf{T}-\mathbf{x}\mathbf{B}) \rbrace$ and $\int R(\mathbf{x}) d\mathbf{x} < \infty$.
	\noindent

	Assumption (A1) assumes that the minimum signal cannot be too small, which is typically used in the feature screening literature (\citet{fan2008sure}). Assumption (A2) is also a typical assumption of population ranking consistency, which ensures that the active covariates are consistently ranked on the top with an overwhelming probability. Under Assumptions (A1) and (A2), one can show that the conditional independence screening procedure based on the generalised covariance measure statistic enjoys the sure screening property and the ranking consistency property. Assumptions (A3) requires the convergence rate of the mean squared prediction error of $\hat{f}$ and $\hat{g}$, which can be satisfied when $f$ and $g$ being in a reproducing kernel Hillbert space (RKHS) and they can be estimated by kernel ridge regression (\citet{shah2020hardness}). Assumption (A4) requires the finite second moment of $\mathbf{X}$.
	
\textbf{Theorem 1}:\\
		For any $0 < \eta \leq 1/2$, if $p$ satisfies that $p \cdot \text{exp}\lbrace -an^{1-2\eta}\rbrace \to 0$ for some positive constant $a$, then 
		\begin{enumerate}
			\item[(\romannumeral 1)] Under Assumption (A1) and (A3), 
			\begin{equation}
				\mathbb{P}(\text{max}_{1\leq k \leq p} \mid \hat{GCM}_k-GCM_k \mid > cn^{-\eta}) \leq O(p \ \text{exp}\lbrace -an^{1-2\eta}\rbrace)
			\end{equation}
			\item[(\romannumeral 2)] Under Assumption (A1) and (A3), we have
			\begin{equation}
				\mathbb{P}(\mathcal{M}^*_{C\cup P} \subseteq \mathcal{M}_{C\cup P}) \geq 1- O(\mid \mathcal{M}^*_{C\cup P}\mid \text{exp}\lbrace -an^{1-2\eta}\rbrace).
			\end{equation}
			\item[(\romannumeral 3)] Under Assumptions (A1)- (A3), it follows that
			\begin{equation}
				\mathbb{P}(\text{lim inf}_{n \to \infty} \lbrace \text{min}_{k \in \mathcal{A}} \hat{GCM}_k - \text{max}_{k \in \mathcal{A}^c} \hat{GCM}_k \rbrace >0) = 1.
			\end{equation}
		\end{enumerate}

	The sure screening property of the conditional independence screening procedure is claimed in Theorem 1(\romannumeral 2), which ensures that the probability of including all the truly active covariates goes to one at an exponential rate as $n \to \infty$. Moreover, the ranking consistency in Theorem 1(\romannumeral 3) guarantees that the conditional independence screening procedure can rank the active covariates above none-active ones with probability one.
	
	Next, theoretical guarantees for the prior adaptive Lasso procedure (Step 3) are provided.
	
\textbf{Theorem 2}:
		Suppose $\lambda n^{1/2} \to 0$, $\lambda n^{\gamma/2-1} \to \infty$ for some $\gamma >3$ as $n \to \infty$ and Assumption (A1)-(A4) holds, then $P(\hat{B}_{ij}=0, \forall i \in \mathcal{A}^c) \to 1$ and other elements of $\hat{B}$ and $\hat{\Theta}$ have the same limiting distribution as those of the maximum likelihood estimates based on the true treatment assignment model.

	Theorem 2 shows that when the tuning parameters are chosen properly, the prior adaptive lasso method can force the estimates $\hat{B}_{ij}$ corresponding to instrumental variables and spurious variables to zero.
	\noindent
	
	The proofs of Theorem 1 and Theorem 2 can be found in the supplemetary materials.
	
	\section{Numerical Simulation\label{sec:5}}
	
	In this section, the performance of the DSPAL method is evaluated and compared to other methods under various simulation scenarios.
	
	\subsection{Data Generating Process}
	\noindent
	
	The simulation mainly focuses on two-dimensional treatments, i.e. $q=2$. Four different combinations of sample size $n$ and covariate dimension $p$ are considered: 
	\begin{equation*}
		(n,p) = (300,100), (500,200), (300,800),(500,1500).
	\end{equation*} 
	Let the covariates be independently drawn from $N_p(\mathbf{0}, I_p)$ and the error matrix $\mathbf{E} \sim N_q(\mathbf{0},\mathbf{\Sigma}_E)$, where the $(i,j)$th element of $\mathbf{\Sigma}_E = (m_{ij})$ is given by $m_{ij} = 0.5^{\mid i-j \mid}$. Let $X_C= (X_1,\dots, X_{5})$ be confounders which associate with the outcome and at least one treatment; $X_P =(X_{6},\dots,X_{10})$ are predictors which associate with the outcome but not with treatments; $X_I=(X_{11},\dots,X_{15})$ are instrumental variables which associate with at least one treatment but not with the outcome; the remaining covariates are spurious variables ($X_S$) which are independent of both the outcome and treatments. 
	
	Both linear and nonlinear outcome generating processes are considered. For the linear outcome model, let $Y_i = \sum_{j=1}^{10}X_{ij}+T_{i1}+T_{i2}+\epsilon_i$, where $\epsilon_i \sim N(0,1)$. For the nonlinear outcome model, let $Y_i = \sum_{j=1}^{10}X_{ij}+T_{i1}+T_{i2}+T_{i1}^2+T_{i2}^2+0.2T_{i1}T_{i2}+\epsilon_i$, where $\epsilon_i \sim N(0,1)$. Since $X_C$ and $X_I$ are associated with treatments while $X_P$ and $X_S$ are not associated with treatments, the rows of $\mathbf{B}$ corresponding to $X_C$ and $X_I$ should include non-zero entries. To ensure that, let the first column of the coefficient matrix $\mathbf{B}$ be $(\underbrace{1,\dots,1}_{5},\underbrace{0,\dots,0}_{5},\underbrace{ 1,\dots,1}_{5},\mathbf{0})^{'}$, the elements of the second column corresponding to the counfounders and instrumental variables follow a binomial distribution with probability 0.5. 
	
	BIC criteria is used to select optimal $\lambda$ over a set of values: \begin{equation*}
		\lbrace n^{-1}, n^{-1.2}, n^{-1.4}, n^{-1.6},n^{-1.8},n^{-2} \rbrace
	\end{equation*}
	for each data setting. $\gamma$ is determined by $\lambda n^{\gamma/2-1} = n^{0.2}$ for each $\lambda$ value. Besides, set D=20 and $K=\lfloor n/\text{log}(n) \rfloor$. 500 independent simulation experiments are run for each numerical setting. 
	
	\subsection{Simulation Results}
	\noindent
	
	The proposed DSPAL method (four steps) is applied to the four cases of combinations of $n$ and $p$, the results that evaluate its overall performance are shown in Tables 1-2 and Figures 2-3. However, it is hard to compare our results with other methods in the case of high dimensional covariates since there are no other methods that deal with this. Hence we apply the first two screening steps to the simulated data, and compare the third step of our proposed method DSPAL with other methods in terms of the variable selection performance for low-dimensional covariates. 
	
	Specifically, DSPAL is compared with ACPME and PALUT based on the covariate set obtained from the conditional independence screening procedure (Step 2), where ACPME refers to the method proposed by \citet{wilson2018model}, which selects covariates by using Bayesian model averaging based on linear outcome model with multivariate treatment, and PALUT refers to the method that selects covariates for each treatment seperately by using the same adaptive lasso method as DSPAL method, and takes the union of the selected covariates as the target covariate set. Since the covariates are high-dimensional, only the proportions of being selected among the 500 independent experiments for the first fifty covariates are shown, the remaining covariates all belong to $X_S$, whose proportions of being selected are similar among different methods.
	\noindent
	
	Figure 2 shows the results of covariate selection in the case of linear outcome model. Observe that DSPAL performs the best among the three methods. Specifically, all three methods select confounders and predictors with high proportions, DSPAL and ACPME even select confounders with proportion one in all scenarios. However, ACPME selects all instrumental variables with proportion one, PALUT selects more spurious variables than DSPAL and ACPME and it selects more instrumental variables than DSPAL. 
	\noindent
	
	\begin{figure}[t!]
		\centering
		\subfigure{
			\begin{minipage}{12cm}
				\centering
				\includegraphics[width=12cm]{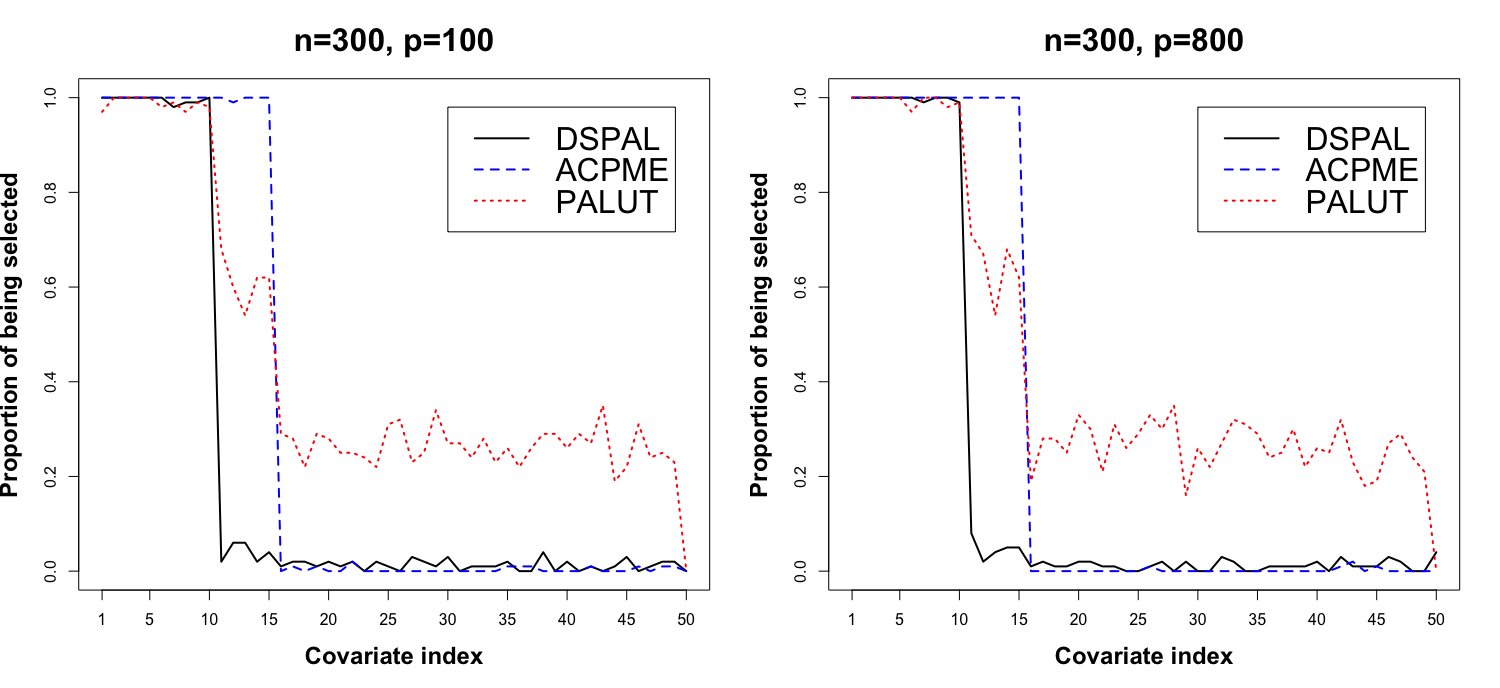} 
			\end{minipage}
		}
		\setcounter{subfigure}{0}
		\subfigure{
			\begin{minipage}{12cm}
				\centering
				\includegraphics[width=12cm]{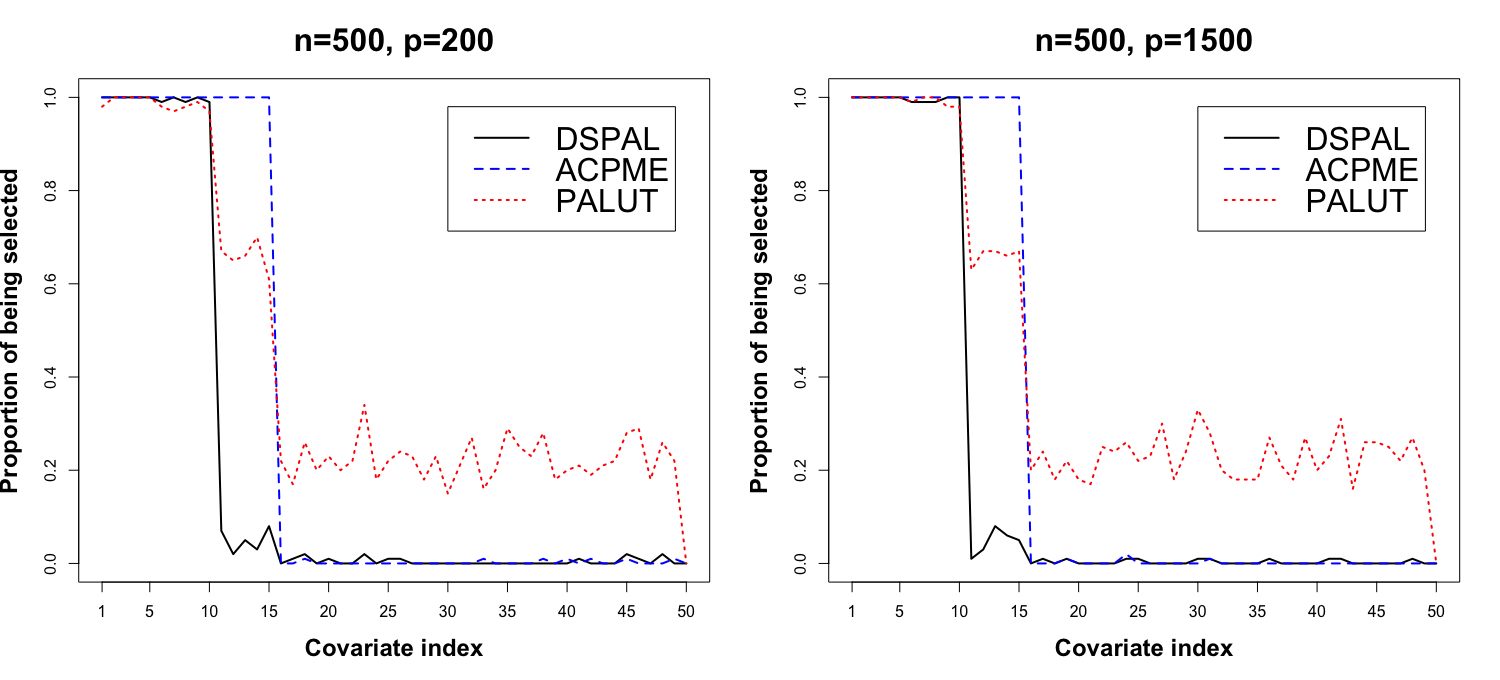} 
			\end{minipage}
			
		}
		\setcounter{subfigure}{0}
		\caption{Performance comparison of covariate selection for the linear outcome model. The covariate indices of 1-5 represent $X_C$, 6-10 represent $X_P$, 11-15 represent $X_I$ and the remaining represent $X_S$. }
		\label{Fig2}
	\end{figure}

	Figure 3 shows the results of covariate selection in the case of nonlinear outcome model. Observe that DSPAL also performs the best among the three methods. DSPAL selects all confounders with proportion one and all predictors with a high proportion. PALUT also selects confounders and predictors with a high proportion but it selects more spurious variables than DSPAL and ACPME. Among the three methods, ACPME performs the worst in that it selects all confounders and instrumental variables with proportion one and predictors with a low proportion. This phenomenon is reasonable since DSPAL and PALUT are suitable for nonparametric outcome model while ACPME assumes a linear outcome model. This further demonstrates that DSPAL is robust to different forms of the outcome model.
	\noindent
	\begin{figure}[t!]
		\centering
		\subfigure{
			\begin{minipage}{12cm}
				\centering
				\includegraphics[width=12cm]{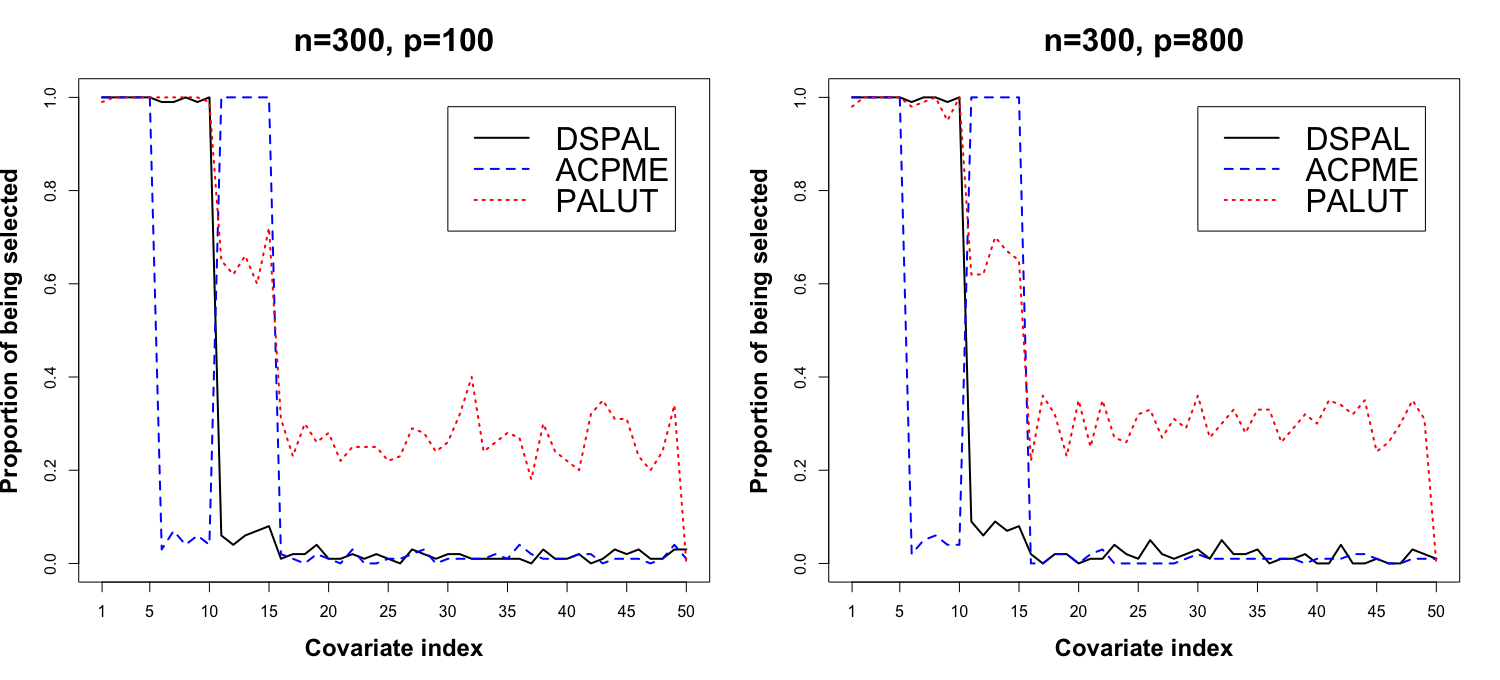} 
			\end{minipage}
		}
		\setcounter{subfigure}{0}
		\subfigure{
			\begin{minipage}{12cm}
				\centering
				\includegraphics[width=12cm]{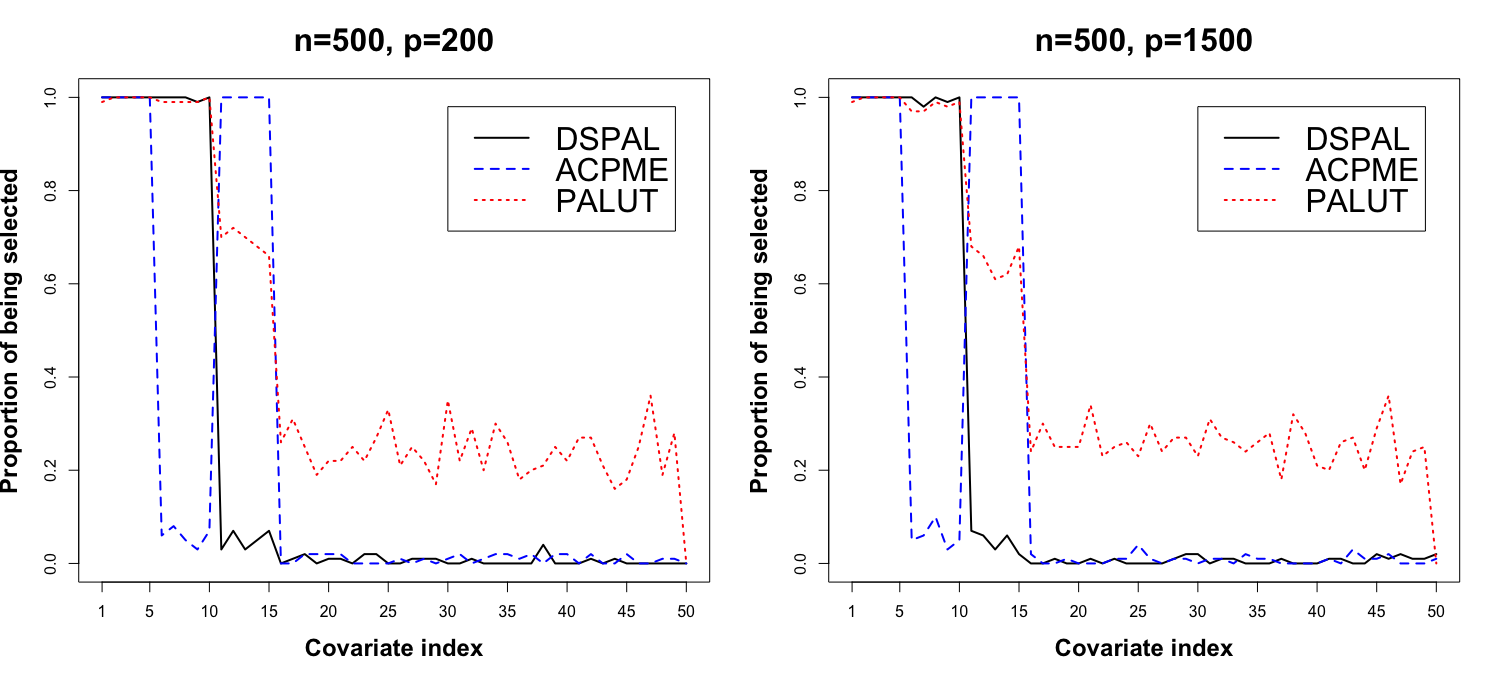} 
			\end{minipage}
		}
		\setcounter{subfigure}{0}
		\caption{Performance comparison of covariate selection for the nonlinear outcome model. The covariate indices of 1-5 represent $X_C$, 6-10 represent $X_P$, 11-15 represent $X_I$ and the remaining represent $X_S$. }
		\label{Fig3}
	\end{figure}
	
	Table 1 shows the results of causal effect estimation for the linear outcome model. It can be seen that the mean bias and RMSE of the estimated coefficients based on DSPAL is the smallest in all screnarios. ACPME method performs the worst as it includes all instrumental variables. Literatures have shown that inclusion of such variables can cause additional bias and inflated variance (\citet{de2011covariate}, \citet{patrick2011implications}). In addition, for all three methods, the mean bias and RMSE decrease as sample size increases and incease as the number of covariates increases.
	\noindent
	\begin{table}[t!]
		\begin{center}
			\caption{Mean bias and RMSE of each regression coefficient of the linear outcome model.}
			\label{tableone}
		\begin{tabular}{cccccc}
					\hline
					\multirow{2}*{$(n,p)$}&\multirow{2}*{	\textbf{Method} } & \multicolumn{2}{c}{\textbf{$\hat{\beta}_1$ ($\beta_1^* = 1$) }} & \multicolumn{2}{c}{ \textbf{$\hat{\beta}_2$ ($\beta_2^* = 1$) }}\\
					\cline{3-6}
					~ &~& Mean Bias & RMSE & Mean Bias & RMSE\\ \hline
					\multirow{3}*{(300,100)}&
					DSPAL& \textbf{0.019} &\textbf{0.072}& \textbf{0.032} & \textbf{0.087} \\ 
					~&PALUT&-0.061& 0.281& 0.128& 0.385\\ 
					~&ACPME& 0.309& 0.523& 0.194 & 0.630 \\ \hline
					\multirow{3}*{(300,800)}&
					DSPAL& \textbf{0.026} &\textbf{0.079}& \textbf{0.048} & \textbf{0.176} \\ 
					~&PALUT&0.170& 0.449& -0.139& 0.409\\ 
					~&ACPME& 0.341& 0.642& -0.211 & 0.697\\ \hline
					\multirow{3}*{(500,200)}&
					DSPAL& \textbf{0.010} &\textbf{0.048}& \textbf{0.024}& \textbf{0.061}\\ 
					~&PALUT&0.041& 0.092& 0.045& 0.099\\ 
					~&ACPME& -0.096& 0.401& 0.138 & 0.493 \\ \hline
					\multirow{3}*{(500,1500)}&
					DSPAL& \textbf{0.022} &\textbf{0.063}&\textbf{ -0.027} & \textbf{0.081} \\ 
					~&PALUT&0.044& 0.196& -0.114& 0.270\\ 
					~&ACPME& 0.222& 0.427& -0.158& 0.503 \\ \hline
					
			\end{tabular}
		\end{center}
	\end{table}
	
	Table 2 shows the results of causal effect estimation for the nonlinear outcome model. Observe that the mean RMSE of DSPAL is the smallest and that of ACPME is the largest since ACPME includes all instrumental variables and almost excludes all predictors. Similarly, the mean RMSE of all methods decreases as sample size increases and increases as the number of covariates increases.

	\begin{table}[t!]
		\begin{center}
			\caption{Mean RMSE of estimated causal effect function for nonlinear outcome model.}
			\label{tabletwo}
			\begin{tabular}{ccccc}
					\hline
					\multirow{2}*{\textbf{Method} }& \multicolumn{4}{c}{Mean RMSE under different $(n,p)$}\\
					\cline{2-5}
					~&(300,100)&(300,800)&(500,200)&(500,1500) \\ \hline
					DSPAL& 2.658&2.825&2.472&2.442  \\ 
					PALUT&11.671&13.874&4.399&6.998\\ 
					ACPME& 36.546&47.136&23.179&33.918 \\  \hline
			\end{tabular}
		\end{center}
	\end{table}

	\section{Data Analysis\label{sec:6}}
	
	The proposed method is applied to finding causal relationship between environmental factors and serum lipid levels. Serum lipid levels are risk factors affecting diseases such as coronary heart disease, Type 2 diabetes and stroke. Both genetic and environmental factors influence the phenotype of lipid levels. Literatures have shown that genetics can affect the lipid level variation (\citet{1993Genetic}, \citet{2005Relative}). Patel et al. (2012) proposed the Environment-wide Association Study (EWAS) method to study the influence of environmental factors on serum lipid levels, utilizing the National Health and Nutrition Examination Survey (NHANES), a public data source. In an EWAS analysis, Patel et al. (2012) selected important environmental factors independently from their marginal association with different serum lipid levels. Besides, they binned all environmental factors into different groups using a categorization provided by NHANES. Based on these groups,  \citet{antonelli2020causal} studied these data using a group-level analyses. The exposure level defined in their analysis is the average level across all environmental factors within the same group, which leads to a univariate continuous treatment. However, it might lose important information to consider a univariate treatment rather than a multivariate treatment, hence we apply the DSPAL method to investigate the causal relationship between multivariate environmental treatments and serum lipid levels.
	\noindent
	
	Literatures have shown that higher exposure to urinary phenolic compounds is linked to obesity and associated with adverse metabolic health outcomes, such as Type 2 diabetes and cardiovascular disease (CVD) (\citet{2015Endocrine}, \citet{2019Environmental}). However, all these researches focus on the association relationship, while one may be intertested in their causal relationship. Therefore, the DSPAL method is applied to the same data of the urinary phenols group used in \citet{antonelli2020causal}, to investigate the causal effects. The data set contains 179 samples,  81 potential confounders and 3 treatments: Urinary 4-tert-octyl phenol (URX4TO), Urinary Benzophenone-3 (URXBP3) and Urinary Bisphenol A (URXBPH). Low-density lipoprotein-cholesterol (LDL) is considered as the outcome, which is a typical index of serum lipid levels.  
	\noindent
	
	The three methods (DSPAL, ACPME, PALUT) are applied to select confounders and predictors. The numbers of selected covariates are 6, 12, and 40, respectively. The six common variables selected by the three methods are PCB156, PCB180, PCB157, Blood Ethylbenzene, Blood o-Xylene and Triceps Skinfold, which have previously been found to be strongly associated with LDL (\citet{ljunggren2014persistent}, \citet{vitali2006exposure}, \citet{terry1989regional}, \citet{goldberg1984changes}). Note that DSPAL didn't select any other covariates which might belong to $X_I$ or $X_S$, and this will be demonstrated by its estimation accuracy shown later. The EBMT method is used to estimate the causal effect function based on the selected covariates. The covariate balancing statistic is $3.018e^{-9}$ for DSPAL method, $7.403e^{-7}$ for ACPME method and 1.894 for PALUT method, which implies the DSPAL and ACPME methods perform well. 
	\noindent


	A linear outcome model is assumed in this analysis and the causal effect estimates are obtained by conducting a weighted linear regression between the three dimensional treatments and LDL. For each of the three methods (DSPAL, ACPME, PALUT), the bootstrap method is used to obtain the standard error and confidence interval of the estimates based on 500 bootstrap replicates.
	
	\noindent
	
	Table 3 shows the estimated causal effects of the URX4TO, URXBP3 and  URXBPH on LDL as well as their standard errors and confidence intervals. It can be seen that ACPME and PALUT didn't find any factors that have significant causal effect since all their confidence intervals contain zero. However, DSPAL found that URXBPH increases LDL significantly, which demonstrates the causal relationship between URXBPH and LDL. Note that URXBPH has previously been found to be significantly associated with LDL (\citet{2022Pregnancy}, \citet{oguazu2017bisphenol}). Furthermore, among the three methods, the standard errors and confidence intervals obtained by DSPAL is the smallest. These all imply that DSPAL selected the best subset of covariates and thus its estimation accuracy is the highest.
	
	\begin{table}[t!]
		\begin{center}
			\caption{Causal effect estimation of the three dimensional treatment (URX4TO, URXBP3, URXBPH) on the outcome (LDL). }

		\begin{tabular}{ccccc}
					
					\hline
					Treatment &Method &Estimate& Standard Error& 95$\%$ CI \\ \hline
					\multirow{3}*{URX4TO}&
					DSPAL& 0.001&0.050& (-0.063,0.121) \\ 
					~&ACPME &0.009& 0.083& (-0.189,0.147)\\ 
					~&PALUT & -0.057& 0.058 & (-0.127,0.114)  \\ 
					
					\hline
					\multirow{3}*{URXBP3}&DSPAL& -0.001&0.002&(-0.002,0.003) \\ 
					~&ACPME &-0.001&0.003&(-0.004,0.007)\\ 
					~&PALUT & 0.001&0.002&(-0.002,0.004)  \\ 
					
					\hline
					\multirow{3}*{URXBPH}&DSPAL& 0.374& 0.329&(0.102,0.952)\\ 
					~&ACPME &	0.028& 0.627& (-1.755,0.738)\\ 
					~&PALUT & -0.667& 0.362 & (-0.657,0.591)  \\ \hline
					
			\end{tabular}
		\end{center}
	\end{table}

	\section{Conclusion and Discussion\label{sec:7}}
	In this paper, a new method called double screening prior adaptive Lasso (DSPAL) is proposed to estimate causal effect of multivariate continuous treatments with high or even ultra-high dimensional covariates. The DSPAL method selects confounders and predictors by combining the usual adaptive Lasso method with the marginal conditional (in)dependence prior information. It can be used in the case of both parametric and nonparametric outcome model, which makes it more robust compared to other methods. Simulation results show that the proposed method outperforms other methods under different scenarios. Finally, the proposed method is applied to investigate the causal relationship between the urinary phenols and low-density lipoprotein-cholesterol (LDL), the result shows that Urinary Bisphenol A (URXBPH) increases LDL significantly.
	\noindent
	
	Future work will be carried out from the following aspects. First, the proposed method assumes a parametric linear Gaussian treatment assignment model, however, one can relax the Gaussian assumption by minimizing discrepancy of $\mathbf{T}$ and $\mathbf{X}\mathbf{B}$, and the screening canonical correlation statistic may be replaced by other correlation statistic such as rank correlation statistic. Besides, the proposed method assumes the target covariate set $X_C \cup X_P$ to be low-dimensional. One extension of this work is to consider high-dimensional target confounders ($X_C$) and predictors ($X_P$). 
	\noindent
	
	Fast development of technology brings opportunity to collect large amounts of data, which makes high-dimensional covariates more and more common. Besides, there is growing interest in investigating the causal relationship between multivariate treatments on outcomes such as complex diseases  (\citet{wild2005complementing}). The main challenge is to select appropriate covariates for estimating the effects of multivariate treatments. The proposed method fills this gap and performs favorably with other methods.

	\section*{Acknowledgments}
	
	The work was supported by National Natural Science Foundation of China (project number: 11771146, 11831008), the 111 Project (B14019) and Program of Shanghai Subject Chief Scientist (14XD1401600). 
	
	\section*{Appendix}
	\subsection*{Proof of the updating formula for $B_{ij}$ }
	
	This section provides the proof of equation (3.16) referred in Section 3.3.
	\noindent
	
	Let 
	\begin{equation}
		g(\mathbf{B}) = \text{tr}[\frac{1}{n}(\mathbf{T}-\mathbf{X}\mathbf{B})^{'}(\mathbf{T}-\mathbf{X}\mathbf{B})\mathbf{\Omega}]+\lambda \sum_{i=1}^{p}\sum_{j=1}^{q}\hat{w}_{ij}\mid B_{ij} \mid,
	\end{equation}
	then 
	\begin{equation}
		\frac{\partial g(\mathbf{B})}{\partial B_{ij}} = \frac{2}{n}\mathbf{e}_j^{'}(\mathbf{X}^{'}\mathbf{X}\mathbf{B}\mathbf{\Omega})\mathbf{e}_i + \lambda \hat{w}_{ij} sgn(B_{ij})- \frac{2}{n}\mathbf{e}_j^{'}(\mathbf{X}^{'}\mathbf{T}\mathbf{\Omega})\mathbf{e}_i,
	\end{equation}
	where $\mathbf{e}_j$ and $\mathbf{e}_i$ are the corresponding base vectors with $p$ and $q$ dimensions. Setting the equation (S1.2) to zero, we have
	\begin{equation}
		\begin{split}
			\frac{\partial g(\mathbf{B})}{\partial B_{ij}} &= \frac{2}{n}\mathbf{e}_j^{'}(\mathbf{X}^{'}\mathbf{X}\mathbf{B}\mathbf{\Omega})\mathbf{e}_i + \lambda \hat{w}_{ij} sgn(B_{ij})- \frac{2}{n}\mathbf{e}_j^{'}(\mathbf{X}^{'}\mathbf{T}\mathbf{\Omega})\mathbf{e}_i +\frac{2}{n}(\mathbf{e}_j^{'}\mathbf{X}^{'}\mathbf{X}\mathbf{e}_j)(\mathbf{e}_i^{'}\mathbf{\Omega}\mathbf{e}_i) B_{ij} \\
			&= \frac{2}{n}(\mathbf{e}_j^{'}\mathbf{X}^{'}\mathbf{X}\mathbf{e}_j)(\mathbf{e}_i^{'}\mathbf{\Omega}\mathbf{e}_i) B_{ij}. 
		\end{split}
	\end{equation}
	Let
	\begin{equation*}
		h_{ij} = -\frac{2}{n}\mathbf{e}_j^{'}(\mathbf{X}^{'}\mathbf{X}\mathbf{B}\mathbf{\Omega})\mathbf{e}_i + \frac{2}{n}\mathbf{e}_j^{'}(\mathbf{X}^{'}\mathbf{T}\mathbf{\Omega})\mathbf{e}_i +\frac{2}{n}(\mathbf{e}_j^{'}\mathbf{X}^{'}\mathbf{X}\mathbf{e}_j)(\mathbf{e}_i^{'}\mathbf{\Omega}\mathbf{e}_i) B_{ij},
	\end{equation*}
	then
	\begin{equation}
		\lambda \hat{w}_{ij} sgn(B_{ij})-h_{ij}+\frac{2}{n}(\mathbf{e}_j^{'}\mathbf{X}^{'}\mathbf{X}\mathbf{e}_j)(\mathbf{e}_i^{'}\mathbf{\Omega}\mathbf{e}_i) B_{ij} = 0,
	\end{equation}
	which implies that 
	\begin{equation}
		B_{ij} = \frac{n(h_{ij}- \lambda \hat{w}_{ij} sgn(B_{ij}))}{2(\mathbf{e}_j^{'}\mathbf{X}^{'}\mathbf{X}\mathbf{e}_j)(\mathbf{e}_i^{'}\mathbf{\Omega}\mathbf{e}_i)}.
	\end{equation}
	Then it can be discussed in the following cases.
	\noindent
	
	If $h_{ij} >  \lambda \hat{w}_{ij}$, then $B_{ij}>0 $ since $sgn(B_{ij}) \leq 1$, and we can obtain that 
	\begin{equation*}
		B_{ij} = \frac{n(h_{ij}- \lambda \hat{w}_{ij}) }{2(\mathbf{e}_j^{'}\mathbf{X}^{'}\mathbf{X}\mathbf{e}_j)(\mathbf{e}_i^{'}\mathbf{\Omega}\mathbf{e}_i)}.
	\end{equation*}
	If $h_{ij} <-  \lambda \hat{w}_{ij}$, then $B_{ij}<0 $ since $sgn(B_{ij}) \geq -1$, and we can obtain that 
	\begin{equation*}
		B_{ij} = \frac{n(h_{ij}+ \lambda \hat{w}_{ij}) }{2(\mathbf{e}_j^{'}\mathbf{X}^{'}\mathbf{X}\mathbf{e}_j)(\mathbf{e}_i^{'}\mathbf{\Omega}\mathbf{e}_i)}.
	\end{equation*}
	If $-  \lambda \hat{w}_{ij} \leq h_{ij} \leq  \lambda \hat{w}_{ij}$, suppose $B_{ij} >0$, then 
	\begin{equation*}
		B_{ij} = \frac{n(h_{ij}- \lambda \hat{w}_{ij}) }{2(\mathbf{e}_j^{'}\mathbf{X}^{'}\mathbf{X}\mathbf{e}_j)(\mathbf{e}_i^{'}\mathbf{\Omega}\mathbf{e}_i)} <0,
	\end{equation*}
	which contradicts with $B_{ij}>0$.
	
	\noindent
	Similarly, suppose $B_{ij} <0$, then 
	\begin{equation*}
		B_{ij} = \frac{n(h_{ij}+ \lambda \hat{w}_{ij}) }{2(\mathbf{e}_j^{'}\mathbf{X}^{'}\mathbf{X}\mathbf{e}_j)(\mathbf{e}_i^{'}\mathbf{\Omega}\mathbf{e}_i)} >0,
	\end{equation*}
	which contradicts with $B_{ij}<0$. Therefore, $B_{ij}=0$ if $-  \lambda \hat{w}_{ij} \leq h_{ij} \leq  \lambda \hat{w}_{ij}$.
	\noindent
	
	Combining these discussions together, we can obtain that 
	\begin{equation*}
		B_{ij} = sgn(h_{ij}) \frac{n(\mid h_{ij} \mid-\lambda \hat{w}_{ij})_+}{2(\mathbf{e}_j^{'}\mathbf{X}^{'}\mathbf{X}\mathbf{e}_j)(\mathbf{e}_i^{'}\mathbf{\Omega}\mathbf{e}_i)},
	\end{equation*}
	hence the proof of the updating formula for $B_{ij}$ is completed.
	
	\subsection*{Proof of bi-convexity}
	This section provides the proof of bi-convexity of the optimization equation (3.12) referred in Section 3.3.
	\noindent
	
	Let
	\begin{equation*}
		L(\textbf{B},\mathbf{\Omega})= tr\lbrace \frac{1}{n}(\textbf{T}-\textbf{X}\textbf{B})^{'}(\textbf{T}-\textbf{X}\textbf{B})\mathbf{\Omega} \rbrace-log\mid \mathbf{\Omega} \mid,
	\end{equation*}
	then one can show that $	L(\textbf{B},\mathbf{\Omega})$ is a bi-convex function of $\textbf{B}$ and $\mathbf{\Omega}$, which means that $	L(\textbf{B},\mathbf{\Omega})$ is a convex function of $\mathbf{B}$ for any fixed $\mathbf{\Omega}$, and $L(\textbf{B},\mathbf{\Omega})$ is a convex function of $\mathbf{\Omega}$ for any fixed $\mathbf{B}$.
	\noindent
	
	The first derivative of $L(\textbf{B},\mathbf{\Omega})$ is 
	\begin{equation*}
		\begin{split}
			dL(\textbf{B},\mathbf{\Omega})&= - tr(\mathbf{\Omega}^{-1}d\mathbf{\Omega}) + tr(\frac{1}{n}\mathbf{T}^{'}\mathbf{T}d\mathbf{\Omega})-2tr(\frac{1}{n}\mathbf{T}^{'}\mathbf{X}\mathbf{\Omega}d\mathbf{B})-2tr(\frac{1}{n}\mathbf{X}^{'}\mathbf{T}\mathbf{B}^{'}d\mathbf{\Omega})\\
			&+2tr(\frac{1}{n}\mathbf{X}^{'}\mathbf{X}\mathbf{B}^{'}\mathbf{\Omega}d\mathbf{B})+tr(\frac{1}{n}\mathbf{B}^{'}\mathbf{X}^{'}\mathbf{X}\mathbf{B}d\mathbf{\Omega})
		\end{split}
	\end{equation*}
	Let $H_T = \frac{1}{n}\mathbf{T}^{'}\mathbf{T}, H_{TX}= \frac{1}{n}\mathbf{X}^{'}\mathbf{T}, H_X= \frac{1}{n}\mathbf{X}^{'}\mathbf{X}$, then 
	\begin{equation*}
		\begin{split}
			dL(\textbf{B},\mathbf{\Omega})&= - tr(\mathbf{\Omega}^{-1}d\mathbf{\Omega}) + tr(H_T\mathbf{T}d\mathbf{\Omega})-2tr(H_{TX}^{'}\mathbf{\Omega}d\mathbf{B})-2tr(H_{TX}\mathbf{B}^{'}d\mathbf{\Omega})\\
			&+2tr(H_X\mathbf{B}^{'}\mathbf{\Omega}d\mathbf{B})+tr(\mathbf{B}^{'}H_X\mathbf{B}d\mathbf{\Omega}).
		\end{split}
	\end{equation*}
	The second derivative of $L(\textbf{B},\mathbf{\Omega})$ is 
	\begin{equation*}
		\begin{split}
			d^{2}L(\textbf{B},\mathbf{\Omega})&=  tr(\mathbf{\Omega}^{-1}d\mathbf{\Omega}\mathbf{\Omega}^{-1}d\mathbf{\Omega}) -4tr(H_{TX}^{'}d\mathbf{\Omega}d\mathbf{B})\\
			&+2tr(H_Xd\mathbf{B}^{'}\mathbf{\Omega}d\mathbf{B})+3tr(H_X\mathbf{B}^{'}d\mathbf{\Omega}d\mathbf{B})+tr(\mathbf{B}H_Xd\mathbf{B}^{'}d\mathbf{\Omega}).
		\end{split}
	\end{equation*}
	Let
	\begin{equation*}
		vec(\textbf{B},\mathbf{\Omega})= \begin{pmatrix} vec(\mathbf{\Omega})\\vec(\mathbf{B}) \end{pmatrix},
	\end{equation*}
	since $tr(ABCD)=(vecB^{'})^{'}(A^{'}\otimes C)(vec(D))$ and $tr(H_X\mathbf{B}^{'}d\mathbf{\Omega}d\mathbf{B}) = tr(\mathbf{B}H_Xd\mathbf{B}^{'}d\mathbf{\Omega})$, we have
	\begin{equation*}
		\begin{split}
			&tr(\mathbf{\Omega}^{-1}d\mathbf{\Omega}\mathbf{\Omega}^{-1}d\mathbf{\Omega}) = (dvec(\mathbf{\Omega}))^{'}(\mathbf{\Omega}^{-1} \otimes \mathbf{\Omega}^{-1})(dvec(\mathbf{\Omega})),\\
			&tr(H_{TX}^{'}d\mathbf{\Omega}d\mathbf{B}) = (dvec(\mathbf{\Omega}))^{'} (C_{TX} \otimes I_q)(dvec(\mathbf{B})),\\
			&tr(H_Xd\mathbf{B}^{'}\mathbf{\Omega}d\mathbf{B})= (dvec(\mathbf{B}))^{'} (H_X \otimes \mathbf{\Omega})(dvec(\mathbf{\Omega})),\\
			&tr(H_X\mathbf{B}^{'}d\mathbf{\Omega}d\mathbf{B}) = (dvec(\mathbf{\Omega}))^{'} (\mathbf{B}H_X \otimes I_q)(dvec(\mathbf{B})).
		\end{split}
	\end{equation*}
	Hence,
	\begin{equation*}
		\begin{split}
			d^{2}L(\textbf{B},\mathbf{\Omega})&=   (dvec(\mathbf{\Omega}))^{'}(\mathbf{\Omega}^{-1} \otimes \mathbf{\Omega}^{-1})(dvec(\mathbf{\Omega}))-4(dvec(\mathbf{\Omega}))^{'} (C_{TX} \otimes I_q)(dvec(\mathbf{B})) \\
			&+2(dvec(\mathbf{B}))^{'} (H_X \otimes \mathbf{\Omega})(dvec(\mathbf{\Omega}))+4(dvec(\mathbf{\Omega}))^{'} (\mathbf{B}H_X \otimes I_q)(dvec(\mathbf{B})) \\
			&= ((dvec(\mathbf{\Omega}))^{'} ,(dvec(\mathbf{B}))^{'} )^{'} \mathbf{W}\begin{pmatrix} dvec(\mathbf{\Omega})\\dvec\mathbf{B} \end{pmatrix}\\
			&=(dvec(\textbf{B},\mathbf{\Omega}))^{'}\mathbf{W}(dvec(\textbf{B},\mathbf{\Omega})),
		\end{split}
	\end{equation*}
	where 
	\begin{equation*}
		\mathbf{W}= \begin{pmatrix} \mathbf{\Omega}^{-1} \otimes \mathbf{\Omega}^{-1} & -2C_{TX} \otimes I_q+ 2(\mathbf{B}H_X) \otimes I_q \\ -2C_{TX}^{'} \otimes I_q+ 2(H_X\mathbf{B}^{'}) \otimes I_q & 2H_X \otimes \mathbf{\Omega} \end{pmatrix}.
	\end{equation*}
	Therefore, the proof of bi-convexity is completed.
	
	\subsection*{Proof of Theorem 1}
	This section provides the proof of Theorem 1 referred in Section 4. To prove Theorem 1, we need the following lemmas.
	
\textbf{Lemma 1}:
		(Bernstein's inequality (\citet{1996Weak}, Lemma 2.2.9)).  Let $X_1,\dots, X_n$ be independent variables with bounded $[-K,K]$ and zero means. Then for any $\epsilon >0$,
		\begin{equation}
			\mathbb{P}(\mid X_1+\dots+X_n \mid > \epsilon) \leq 2\text{exp}\lbrace -\frac{\epsilon^2}{2(\mu+K\epsilon/3)}\rbrace
		\end{equation}
		for $\mu \geq \text{Var}(X_1+\dots+X_n)$.

	Denote 
	\[GCM^* = \frac{\rho}{\sigma},\]
	where $\rho=\mathbb{E}(\zeta_x\zeta_y), \sigma=\sqrt{\text{var}(\zeta_x\zeta_y)}$.
	
\textbf{Lemma 2}: 
		Under assumption (A1), suppose $\mid R_i \mid \leq M  $ for some positive constant $M$, then for any $\epsilon \in (0,1)$, there exists a constant $b$ such that 
		\begin{equation}
			\mathbb{P}(\mid \hat{GCM}_k-GCM^*_k \mid \geq \epsilon) \leq 8\text{exp}\lbrace-bn\epsilon \rbrace.
		\end{equation}

	Proof of Lemma 2: Denote
	\[\hat{GCM} = \frac{\frac{1}{n}\sum_{i=1}^{n}R_i}{(\frac{1}{n}\sum_{i=1}^{n}R_i^2-(\frac{1}{n}\sum_{k=1}^{n}R_k)^2)^{1/2}} \equiv \frac{\frac{1}{n}\sum_{i=1}^{n}R_i}{s_n},\]
	where $s_n =(\frac{1}{n}\sum_{i=1}^{n}R_i^2-(\frac{1}{n}\sum_{k=1}^{n}R_k)^2)^{1/2}$. Then
	\begin{equation}
		\begin{split}
			\mathbb{P}(\mid \hat{GCM}_k-GCM^*_k \mid \geq 2\epsilon) &= \mathbb{P}(\mid \frac{\frac{1}{n}\sum_{i=1}^{n}R_i}{s_n} - \frac{\rho}{\sigma} \mid > 2\epsilon) \\
			&= \mathbb{P}(\mid \frac{\frac{1}{n}\sum_{i=1}^{n}R_i}{\sigma} \cdot \frac{\sigma}{s_n} -  \frac{\sigma}{s_n} \cdot  \frac{\rho}{\sigma}+\frac{\sigma}{s_n} \cdot  \frac{\rho}{\sigma}- \frac{\rho}{\sigma} \mid > 2\epsilon) \\
			&\leq \mathbb{P}(\mid (\frac{\frac{1}{n}\sum_{i=1}^{n}R_i}{\sigma}- \frac{\rho}{\sigma})\cdot  \frac{\sigma}{s_n} \mid >\epsilon) + \mathbb{P}(\mid\frac{\rho}{\sigma}( \frac{\sigma}{s_n}-1) \mid >\epsilon). 
		\end{split}
	\end{equation}
	Let $W_n =  \frac{\sigma}{s_n}$, $V_n = \frac{\frac{1}{n}\sum_{i=1}^{n}(R_i-\rho)}{\sigma}$, then
	$W_n=1+o_p(1)$(\citet{shah2020hardness}, Theorem 8). Thus,
	\begin{equation}
		\begin{split}
			&\mathbb{P}(\mid (\frac{\frac{1}{n}\sum_{i=1}^{n}R_i}{\sigma}- \frac{\rho}{\sigma})\cdot  \frac{\sigma}{s_n} \mid >\epsilon) + \mathbb{P}(\mid\frac{\rho}{\sigma}( \frac{\sigma}{s_n}-1) \mid >\epsilon) \\
			&= \mathbb{P}(\mid V_n\cdot W_n \mid >\epsilon)+\mathbb{P}(\mid \frac{\rho}{\sigma}(W_n-1) \mid >\epsilon)\\
			&=\mathbb{P}(\mid V_n\cdot W_n -V_n +V_n \mid > \epsilon)+\mathbb{P}(\mid \frac{\rho}{\sigma}(W_n-1) \mid >\epsilon)\\
			&\leq \mathbb{P}(\mid V_n \mid > \frac{\epsilon}{2})+\mathbb{P}(\mid V_n(W_n-1)\mid >\frac{\epsilon}{2})+\mathbb{P}(\mid \frac{\rho}{\sigma}(W_n-1) \mid >\epsilon)\\
			&\leq \mathbb{P}(\mid V_n \mid > \frac{\epsilon}{2})+\mathbb{P}(\mid V_n\mid >\sqrt{\frac{\epsilon}{2}})+\mathbb{P}(\mid W_n-1\mid >\sqrt{\frac{\epsilon}{2}})+\mathbb{P}(\mid \frac{\rho}{\sigma}(W_n-1) \mid >\epsilon)\\
			&\leq 2\mathbb{P}(\mid V_n\mid >\frac{\epsilon}{2})+2\mathbb{P}(\mid W_n-1\mid > c_1\epsilon) \ \ \text{for some $c_1>0$}.
		\end{split}
	\end{equation}
	Since
	\begin{equation}
		\begin{split}
			\mathbb{P}(\mid V_n \mid >\epsilon) &= \mathbb{P}(\mid\sum_{i=1}^{n} \frac{R_i-\rho}{\sigma} \mid > n\epsilon)\\
			&\mathbb{P}(\mid\sum_{i=1}^{n} \frac{R_i-E(R_i)+E(R_i)-\rho}{\sigma} \mid > n\epsilon)\\
			&\leq \mathbb{P}(\mid\sum_{i=1}^{n} \frac{R_i-E(R_i)}{\sigma} \mid > n\epsilon/2)+\mathbb{P}(\mid\sum_{i=1}^{n} \frac{ E(R_i)-\rho}{\sigma} \mid > n\epsilon/2)
		\end{split}
	\end{equation}
	By Lemma 1, we have
	\begin{equation}
		\mathbb{P}(\mid\sum_{i=1}^{n} \frac{R_i-E(R_i)}{\sigma} \mid > n\epsilon/2) \leq 2\text{exp}\lbrace \frac{-n^2\epsilon^2}{8(n+\frac{M n\epsilon}{6\sigma})} \rbrace = 2\text{exp}\lbrace-c_2n\epsilon^2\rbrace.
	\end{equation}
	where $c_2 = \frac{1}{8(1+\frac{M\epsilon}{6\sigma})}$. Since
	\begin{equation}
		\begin{split}
			E(R_i)-\rho &= E(x_i-\hat{f}(z_i))(y_i-\hat{g}(z_i))-\rho\\
			&= E(x_iy_i-x_i\hat{g}(z_i)-y_i\hat{f}(z_i)+\hat{f}(z_i)\hat{g}(z_i))-\rho \\
			& = E(f(z_i)+\zeta_{x,i})(g(z_i)+\zeta_{y,i})-E(f(z_i)+\zeta_{x,i})\hat{g}(z_i)\\
			&-E(g(z_i)+\zeta_{y,i})\hat{f}(z_i)+E\hat{f}(z_i)\hat{g}(z_i))-\rho \\
			&= E((f(z_i)-\hat{f}(z_i))(g(z_i)-\hat{g}(z_i)),
		\end{split}
	\end{equation}
	we have
	\begin{equation}
		\begin{split}
			\mathbb{P}(\mid\sum_{i=1}^{n} \frac{ E(R_i)-\rho}{\sigma} \mid > n\epsilon/2) &= \mathbb{P}(\mid\sum_{i=1}^{n} \frac{ E((f(z_i)-\hat{f}(z_i))(g(z_i)-\hat{g}(z_i))}{\sigma} \mid > n\epsilon/2) \\
			&\leq 	\mathbb{P}(E(\sum_{i=1}^{n}(f(z_i)-\hat{f}(z_i))^2\cdot E(\sum_{i=1}^{n}(g(z_i)-\hat{g}(z_i))^2  > n\epsilon/2) \to 0.
		\end{split}
	\end{equation}
	Similarly, one can show that $\mathbb{P}(\mid W_n-1\mid > \epsilon ) \leq 2\text{exp}\lbrace-c_3n\epsilon^2\rbrace$ for some constant $c_3$, then 
	\begin{equation}
		\mathbb{P}(\mid (\frac{\frac{1}{n}\sum_{i=1}^{n}R_i}{\sigma}- \frac{\rho}{\sigma})\cdot  \frac{\sigma}{s_n} \mid >\epsilon) + \mathbb{P}(\mid\frac{\rho}{\sigma}( \frac{\sigma}{s_n}-1) \mid >\epsilon) 
		\leq 8\text{exp}\lbrace-c_4n\epsilon^2 \rbrace.
	\end{equation}
	Therefore,
	\begin{equation}
		\mathbb{P}(\mid \hat{GCM}_k-GCM^*_k \mid \geq \epsilon) 
		\leq 8\text{exp}\lbrace-c_4n\epsilon^2\rbrace
	\end{equation}
	Hence, the proof of Lemma 2 is completed.
	\\
	\\
	\textbf{Proof of Theorem 1}:
	\\
	By Lemma 2, we have
	\begin{equation}
		\mathbb{P}(\mid \hat{GCM}_k-GCM^*_k \mid \geq cn^{-\eta} ) 
		\leq 8\text{exp}\lbrace-b_1n^{1-2\eta}\rbrace
	\end{equation}
	For Theorem 1(\romannumeral 1), we can get that
	\begin{equation}
		\begin{split}
			&	\mathbb{P}(\mathop{\text{max}} \limits_{1\leq k \leq p}\mid  \hat{GCM}_k-GCM^*_k \mid \geq cn^{-\eta}) \\
			&\leq O(p)\ \mathop{\text{max}} \limits_{1\leq k \leq p}	\mathbb{P}(\mid \hat{GCM}_k-GCM^*_k \mid \geq cn^{-\eta} )\\
			& \leq O(p)\cdot \text{exp}\lbrace-b_1n^{1-2\eta}\rbrace
		\end{split}
	\end{equation}
	For Theorem 1(\romannumeral 2), by assumption (A1), if $GCM_k \in \mathcal{M}^*_{C \cup P} $ but $\hat{GCM}_k \notin \mathcal{M}_{C\cup P}$, then 
	\begin{equation*}
		\mid \hat{GCM}_k-GCM_k \mid > 2cn^{-\eta}-cn^{-\eta}=cn^{-\eta},
	\end{equation*}
	which implies that 
	\begin{equation*}
		\lbrace \mathcal{M}^*_{C \cup P} \nsubseteq \mathcal{M}_{C\cup P} \rbrace \subseteq \lbrace
		\mid \hat{GCM}_k-GCM_k \mid >cn^{-\eta}, \text{for some} \ k \in   \mathcal{M}^*_{C \cup P} \rbrace,
	\end{equation*}
	Thus,
	\begin{equation}
		\begin{split}
			\mathbb{P}(\mathcal{M}^*_{C \cup P} \subseteq \mathcal{M}_{C\cup P} ) &\geq 1- \mathbb{P}\lbrace
			\mid \hat{GCM}_k-GCM_k \mid >cn^{-\eta}, \text{for some} \ k \in   \mathcal{M}^*_{C \cup P} \rbrace \\
			&\geq 1- \mid \mathcal{M}^*_{C \cup P} \mid \mathop{\text{max}} \limits_ {k \in   \mathcal{M}^*_{C \cup P}} \mathbb{P}(\mid \hat{GCM}_k-GCM_k \mid >cn^{-\eta})\\
			&\geq 1-O(\mid \mathcal{M}^*_{C \cup P} \mid \cdot \text{exp}\lbrace-b_1n^{1-2\eta}\rbrace ).
		\end{split}
	\end{equation}
	For Theorem 1(\romannumeral 3), by assumption (A2), we know that there exists some $\delta$ such that 
	\begin{equation*}
		\mathop{\text{min}} \limits_{k \in \mathcal{M}^*_{C \cup P}} GCM_k - \mathop{\text{max}} \limits_{k \in (\mathcal{M}^*_{C \cup P})^c} GCM_k =\delta.
	\end{equation*}
	Then we have 
	\begin{equation*}
		\begin{split}
			&\mathbb{P}\lbrace \mathop{\text{min}} \limits_{k \in \mathcal{M}^*_{C \cup P}} \hat{GCM}_k \leq \mathop{\text{max}} \limits_{k \in (\mathcal{M}^*_{C \cup P})^c} \hat{GCM}_k \rbrace \\
			&= \mathbb{P} \lbrace \mathop{\text{min}} \limits_{k \in \mathcal{M}^*_{C \cup P}} \hat{GCM}_k -\mathop{\text{min}} \limits_{k \in \mathcal{M}^*_{C \cup P}} GCM_k +\delta \leq \mathop{\text{max}} \limits_{k \in (\mathcal{M}^*_{C \cup P})^c} \hat{GCM}_k - \mathop{\text{max}} \limits_{k \in (\mathcal{M}^*_{C \cup P})^c} GCM_k \rbrace \\
			&\leq \mathbb{P} \lbrace \mid (\mathop{\text{min}} \limits_{k \in \mathcal{M}^*_{C \cup P}} \hat{GCM}_k - \mathop{\text{max}} \limits_{k \in (\mathcal{M}^*_{C \cup P})^c} \hat{GCM}_k) - (\mathop{\text{min}} \limits_{k \in \mathcal{M}^*_{C \cup P}} GCM_k 
			- \mathop{\text{max}} \limits_{k \in (\mathcal{M}^*_{C \cup P})^c} GCM_k) \mid \geq \delta \rbrace \\
			&\leq \mathbb{P} \lbrace 2 \mathop{\text{max}} \limits_{1 \leq k \leq p} \mid \hat{GCM}_k - GCM_k \mid \geq \delta \rbrace \\
			&\leq O(p\cdot \text{exp}\lbrace-b_1n^{1-2\eta} \rbrace).
		\end{split}
	\end{equation*}
	By Fatou's Lemma, one can get that
	\begin{equation*}
		\begin{split}
			&	\mathbb{P} \lbrace \mathop{\text{lim inf}} \limits_{n \to \infty} \ (\mathop{\text{min}} \limits_{k \in \mathcal{M}^*_{C \cup P}} \hat{GCM}_k - \mathop{\text{max}} \limits_{k \in (\mathcal{M}^*_{C \cup P})^c} \hat{GCM}_k) \leq 0 \rbrace \\
			&\leq \mathop{\text{lim}} \limits_{n \to \infty} \mathbb{P} (\mathop{\text{min}} \limits_{k \in \mathcal{M}^*_{C \cup P}} \hat{GCM}_k - \mathop{\text{max}} \limits_{k \in (\mathcal{M}^*_{C \cup P})^c} \hat{GCM}_k \leq 0) =0. 
		\end{split}
	\end{equation*}
	Hence
	\begin{equation*}
		\mathbb{P} \lbrace \mathop{\text{lim inf}} \limits_{n \to \infty}\ (\mathop{\text{min}} \limits_{k \in \mathcal{M}^*_{C \cup P}} \hat{GCM}_k - \mathop{\text{max}} \limits_{k \in (\mathcal{M}^*_{C \cup P})^c} \hat{GCM}_k) > 0 \rbrace =1.
	\end{equation*}
	Therefore, the proof of Theorem 1 is completed.
	
	\subsection*{Proof of Theorem 2}
	This section provides the proof of Theorem 2 referred in Section 4. 
	\noindent
	
	Define $\mathcal{A} = C \cup P$, $\mathcal{A}^c = I \cup S$, 
	and
	\begin{equation*}
		\begin{split}
			&\phi_n(U,V) = -\text{log} \mid \mathbf{\Omega} + \frac{U}{\sqrt{n}} \mid \\
			&+ \text{tr} \lbrace (\mathbf{\Omega}+ \frac{U}{\sqrt{n}})(C_Y-C_{YX}(\mathbf{B}+ \frac{V}{\sqrt{n}})^{'}-(\mathbf{B}+ \frac{V}{\sqrt{n}})C_{YX}^{'}+(\mathbf{B}+ \frac{V}{\sqrt{n}})C_X(\mathbf{B}+ \frac{V}{\sqrt{n}})^{'}) \rbrace \\
			&\lambda \sum_{i,j} \hat{w}_{ij} \mid B_{ij}+\frac{V_{ij}}{\sqrt{n}} \mid 
			+\text{log} \mid \mathbf{\Omega} \mid -  \text{tr} \lbrace (\mathbf{\Omega})(C_Y-C_{YX}\mathbf{B}^{'}-\mathbf{B}C_{YX}^{'}+\mathbf{B}C_X\mathbf{B}^{'}) \rbrace   - \lambda \sum_{i,j} \hat{w}_{ij} \mid B_{ij}\mid
		\end{split}
	\end{equation*}
	Using the same argument as in \citet{yuan2007model}, we can show that
	\begin{equation*}
		\text{log} \mid \mathbf{\Omega} + \frac{U}{\sqrt{n}} \mid -\text{log} \mid \mathbf{\Omega} \mid = \frac{1}{\sqrt{n}} \text{tr}(U\Sigma)-\frac{1}{n}\text{tr}(U\Sigma U\Sigma)+o(n^{-1}).
	\end{equation*}	
	Furthermore,
	\begin{equation}
		\begin{split}
			&\text{tr} \lbrace (\mathbf{\Omega}+ \frac{U}{\sqrt{n}})(C_Y-C_{YX}(\mathbf{B}+ \frac{V}{\sqrt{n}})^{'}-(\mathbf{B}+ \frac{V}{\sqrt{n}})C_{YX}^{'}+(\mathbf{B}+ \frac{V}{\sqrt{n}})C_X(\mathbf{B}+ \frac{V}{\sqrt{n}})^{'}) \rbrace\\
			& -\text{tr} \lbrace (\mathbf{\Omega})(C_Y-C_{YX}\mathbf{B}^{'}-\mathbf{B}C_{YX}^{'}+\mathbf{B}C_X\mathbf{B}^{'}) \rbrace \\
			&= \frac{1}{\sqrt{n}} \text{tr} \lbrace \mathbf{\Omega}(VC_X\mathbf{B}^{'}-VC_{YX}^{'}+\mathbf{B}C_XV^{'}-C_{YX}V^{'}) \rbrace\\
			&+\frac{1}{\sqrt{n}} \text{tr} \lbrace U(C_Y-C_{YX}\mathbf{B}^{'}-\mathbf{B}C_{YX}^{'}+\mathbf{B}C_X\mathbf{B}^{'}) \rbrace \\
			&+\frac{1}{n}\text{tr} \lbrace \mathbf{\Omega}(VC_XV^{'}) \rbrace+\frac{1}{n}\text{tr} \lbrace U(VC_X\mathbf{B}^{'}-VC_{YX}^{'}+\mathbf{B}C_XV^{'}-C_{YX}V^{'}) \rbrace \\
			&+\frac{1}{n\sqrt{n}}\text{tr} \lbrace UVC_XV^{'} \rbrace.
		\end{split}
	\end{equation}
	
	Let $P_n = C_{YX}-\mathbf{B}C_X$, then $J_n = \sqrt{n}P_n \to J$, where $J \sim N(0,\Lambda_1)$ with $\text{cov}(J_{ij},J_{i^{'}j^{'}})= \Sigma_{i,i^{'}}(C_X)_{j,j^{'}}$. Similarly, let $Q_n = C_Y-(\Sigma+\mathbf{B}C_X\mathbf{B}^{'})$, then $F_n = \sqrt{n}Q_n \to F$, where $F \sim N(0,\Lambda_2)$ with $\text{cov}(F_{ij},F_{i^{'}j^{'}})=\text{cov}(Y^{(i)}Y^{(j)},Y^{(i^{'})}Y^{(j^{'})} \mid \mathbf{X})$. \\
	
	Then
	\begin{equation*}
		\begin{split}
			(4.16) &=  -\frac{1}{\sqrt{n}} \text{tr} \lbrace \mathbf{\Omega}(VP_n^{'}+P_nV^{'}) \rbrace +\frac{1}{\sqrt{n}} \text{tr} (U\Sigma)
			+\frac{1}{\sqrt{n}} \text{tr} \lbrace U(Q_n-P_n\mathbf{B}^{'}-\mathbf{B}P_n^{'}) \rbrace \\
			&+\frac{1}{n}\text{tr} \lbrace \mathbf{\Omega}(VC_XV^{'}) \rbrace-\frac{1}{n}\text{tr} \lbrace U(VP_n^{'}+P_nV^{'}) \rbrace +\frac{1}{n\sqrt{n}}\text{tr} \lbrace UVC_XV^{'} \rbrace.
		\end{split}
	\end{equation*}
	
	Therefore, we can get that 
	\begin{equation*}
		\begin{split}
			n\phi_n(U,V) &= \text{tr}\lbrace U\Sigma U\Sigma \rbrace + \text{tr}\lbrace \mathbf{\Omega}VC_XV^{'} \rbrace + \text{tr}\lbrace U(F_n-J_n\mathbf{B}^{'}-\mathbf{B}J_n^{'}) \rbrace \\
			&- \text{tr}\lbrace \mathbf{\Omega}(VJ_n^{'}+J_nV^{'}) \rbrace 
			+n\lambda \sum_{i,j} \hat{w}_{ij} (\mid B_{ij}+\frac{V_{ij}}{\sqrt{n}} \mid -\mid B_{ij} \mid)+o(1).
		\end{split}
	\end{equation*}
	
	
	Since for $i \in \mathcal{A}$, we have $\hat{w}_{ij} =O_p(1), \lambda \sqrt{n} \to 0$, then
	\begin{equation*}
		\sqrt{n}\lambda \hat{w}_{ij} \cdot \sqrt{n}(\mid B_{ij}+\frac{V_{ij}}{\sqrt{n}} \mid -\mid B_{ij} \mid) \to 0.
	\end{equation*}
	For $i \in \mathcal{A}^c$, we have $\hat{w}_{ij} = O_p(n^{-\frac{1}{2}}), \lambda n^{\frac{\gamma+1}{2}} \to \infty$, then 
	\begin{equation*}
		\lambda n^{\frac{\gamma+1}{2}}(n^{-\frac{\gamma}{2}}\hat{w}_{ij})\cdot  \sqrt{n}(\mid B_{ij}+\frac{V_{ij}}{\sqrt{n}} \mid -\mid B_{ij} \mid)  \to \infty.
	\end{equation*}
	Therefore, by Slutskey's theorem, we can get that 
	\begin{equation*}
		n\phi_n(U,V) \to \phi(U,V) \ \text{in distribution},
	\end{equation*}
	where 
	\begin{equation*}
		\phi(U,V) = 
		\text{tr}\lbrace U\Sigma U\Sigma \rbrace + \text{tr}\lbrace \mathbf{\Omega}VC_XV^{'} \rbrace + \text{tr}\lbrace  U(F-J\mathbf{B}^{'}-\mathbf{B}J^{'}) \rbrace - \text{tr}\lbrace \mathbf{\Omega}(VJ^{'}+JV^{'}) \rbrace
	\end{equation*}
	for $V$ satisfying that $ V_{ij}=0 \ \text{if}\ i \in \mathcal{A}^c$. And $\phi(U,V) =	\infty$,otherwise.
	
	Since both $\phi(U,V)$ and $n\phi_n(U,V)$ are convex and $\phi(U,V)$ has a unique minimum, we have 
	\begin{equation*}
		\text{argmin}\ n\phi_n(U,V) \to \text{argmin} \ \phi(U,V),
	\end{equation*}
	i.e., 
	\begin{equation*}
		\begin{split}
			\sqrt{n}\lbrace (\hat{\mathbf{\Omega}},\hat{\mathbf{B}}) - (\mathbf{\Omega}, \mathbf{B}) \rbrace \to_d \text{argmin} \ \lbrace &\text{tr}\lbrace U\Sigma U\Sigma \rbrace + \text{tr}\lbrace \mathbf{\Omega}VC_XV^{'} \rbrace \\
			&+ \text{tr}\lbrace U(F-J\mathbf{B}^{'}-\mathbf{B}J^{'}) \rbrace - \text{tr}\lbrace \mathbf{\Omega}(VJ^{'}+JV^{'}) \rbrace,
		\end{split}
	\end{equation*}
	where the minimum is taken over all $p\times q$ matrices $V$ satisfying that $V_{ij} = 0$ for all $i \in \mathcal{A}^c$.
	Hence, we can get that the all elements of $\mathbf{\Omega}$ and the element $\mathbf{B}_{ij}$ for $i \in \mathcal{A}$ have the same limiting distribution as those of the maximum likelihood estimates based on the true treatment assignment model. \\
	
	Next, one can show that $P(\hat{B}_{ij}=0, \forall i \in \mathcal{A}^c) \to 1$.
	Let
	\begin{equation*}
		\begin{split}
			\mathcal{L}_n(\mathbf{B},\mathbf{\Omega}) &= \lbrace\frac{1}{n}(\textbf{T}-\textbf{X}\textbf{B})^{'}(\textbf{T}-\textbf{X}\textbf{B})\mathbf{\Omega} \rbrace-log\mid \mathbf{\Omega}  \mid \notag 
			+ \lambda\sum_{i=1}^{p}\sum_{j=1}^{q}\hat{w}_{ij}\mid B_{ij} \mid 
		\end{split}
	\end{equation*}
	and 
	\begin{equation*}
		L_n(\mathbf{B},\mathbf{\Omega}) = \lbrace \frac{1}{n}(\textbf{T}-\textbf{X}\textbf{B})^{'}(\textbf{T}-\textbf{X}\textbf{B})\mathbf{\Omega} \rbrace-log\mid \mathbf{\Omega}  \mid .
	\end{equation*}
	Denote $D_n(\mathbf{B}_{(\mathcal{A})},\mathbf{B}_{(\mathcal{A}^c)}) \equiv L_n(\mathbf{B}_{(\mathcal{A})},\mathbf{B}_{(\mathcal{A}^c)},\hat{\mathbf{\Omega}})$, then by the mean value theorem, we have 
	\begin{equation*}
		D_n(\mathbf{B}_{(\mathcal{A})},\mathbf{B}_{(\mathcal{A}^c)}) - D_n(\mathbf{B}_{(\mathcal{A})},\mathbf{0}) = \lbrace \frac{\partial  D_n(\mathbf{B}_{(\mathcal{A})},\xi) }{\partial \mathbf{B}_{(\mathcal{A}^c)}}  \rbrace  \mathbf{B}_{(\mathcal{A}^c)}
	\end{equation*}
	for some $\mid \mid \xi \mid \mid \leq \mid \mid \mathbf{B}_{(\mathcal{A}^c)} \mid \mid$. Again, by the mean value theorem, we have
	\begin{equation*}
		\begin{split}
			&	\mid \mid  \frac{\partial  D_n(\mathbf{B}_{(\mathcal{A})},\xi) }{\partial \mathbf{B}_{(\mathcal{A}^c)}} - \frac{\partial  D_n(\mathbf{B}^*_{(\mathcal{A})},\mathbf{0}) }{\partial \mathbf{B}_{(\mathcal{A}^c)}}  \mid \mid \\
			&\leq 	\mid \mid  \frac{\partial  D_n(\mathbf{B}_{(\mathcal{A})},\xi) }{\partial \mathbf{B}_{(\mathcal{A}^c)}} - \frac{\partial  D_n(\mathbf{B}_{(\mathcal{A})},\mathbf{0}) }{\partial \mathbf{B}_{(\mathcal{A}^c)}}  \mid \mid +	\mid \mid  \frac{\partial  D_n(\mathbf{B}_{(\mathcal{A})},\mathbf{0}) }{\partial \mathbf{B}_{(\mathcal{A}^c)}} - \frac{\partial  D_n(\mathbf{B}^*_{(\mathcal{A})},\mathbf{0}) }{\partial \mathbf{B}_{(\mathcal{A}^c)}}  \mid \mid \\
			& \leq \lbrace \sum_{i=1}^{n} R(\mathbf{x}_i) \rbrace \mid \mid \xi \mid \mid + \lbrace \sum_{i=1}^{n} R(\mathbf{x}_i) \rbrace \mid \mid \mathbf{B}_{(\mathcal{A})}- \mathbf{B}^*_{(\mathcal{A})} \mid \mid.
		\end{split}
	\end{equation*}
	For $k \in S$, $ \mid \mid \xi \mid \mid  \leq  \mid \mid \mathbf{B}_S \mid \mid = O_p(n^{-1/2})$. Hence
	\begin{equation*}
		\mid \mid  \frac{\partial  D_n(\mathbf{B}_{(\mathcal{A})},\xi) }{\partial \mathbf{B}_{(\mathcal{A}^c)}} - \frac{\partial  D_n(\mathbf{B}^*_{(\mathcal{A})},\mathbf{0}) }{\partial \mathbf{B}_{(\mathcal{A}^c)}}  \mid \mid  \leq O_p(n^{1/2}).
	\end{equation*}
	For $k \in I$, $ \mid \mid \xi \mid \mid  \leq  \mid \mid \mathbf{B}_I \mid \mid = O_p(1)$. Hence
	\begin{equation*}
		\mid \mid  \frac{\partial  D_n(\mathbf{B}_{(\mathcal{A})},\xi) }{\partial \mathbf{B}_{(\mathcal{A}^c)}} - \frac{\partial  D_n(\mathbf{B}^*_{(\mathcal{A})},\mathbf{0}) }{\partial \mathbf{B}_{(\mathcal{A}^c)}}  \mid \mid  \leq O_p(n).
	\end{equation*}
	Therefore, for $ k \in \mathcal{A}^c$, $	\mid \mid  \frac{\partial  D_n(\mathbf{B}_{(\mathcal{A})},\xi) }{\partial \mathbf{B}_{(\mathcal{A}^c)}} - \frac{\partial  D_n(\mathbf{B}^*_{(\mathcal{A})},\mathbf{0}) }{\partial \mathbf{B}_{(\mathcal{A}^c)}}  \mid \mid  \leq O_p(n) $. \\
	
	Since for fixed $\mathbf{\Omega}$,
	\begin{equation*}
		\begin{split}
			\mathcal{L}_n(\mathbf{B}_{(\mathcal{A})},\mathbf{B}_{(\mathcal{A}^c)}, \mathbf{\Omega}) - 	\mathcal{L}_n(\mathbf{B}_{(\mathcal{A})},\mathbf{0}, \mathbf{\Omega})  &= \sum_{i \in \mathcal{A}^c,j} \lbrace -\mid B_{ij} \mid O_p(n) + \lambda \hat{w}_{ij} \mid B_{ij} \mid \rbrace \\
			&= \sum_{i \in \mathcal{A}^c,j} \lbrace -\mid B_{ij} \mid O_p(n) + \lambda n^{\frac{\gamma}{2}}O_p(1) \mid B_{ij} \mid \rbrace	
		\end{split}
	\end{equation*}
	under the assumption $\lambda n^{\frac{\gamma}{2}-1} \to \infty$, we have   \[\mathcal{L}_n(\mathbf{B}_{(\mathcal{A})},\mathbf{B}_{(\mathcal{A}^c)}, \mathbf{\Omega}) - 	\mathcal{L}_n(\mathbf{B}_{(\mathcal{A})},\mathbf{0}, \mathbf{\Omega})  > 0 \ \text{in probability,}\] which implies that 
	\begin{equation*}
		\begin{split}
			&\mathcal{L}_n(\mathbf{B}_{(\mathcal{A})},\mathbf{B}_{(\mathcal{A}^c)}, \mathbf{\Omega}) - 	\mathcal{L}_n(\hat{\mathbf{B}}_{(\mathcal{A})},\mathbf{0}, \mathbf{\Omega}) \\
			&=  	\mathcal{L}_n(\mathbf{B}_{(\mathcal{A})},\mathbf{B}_{(\mathcal{A}^c)}, \mathbf{\Omega}) - 	\mathcal{L}_n(\mathbf{B}_{(\mathcal{A})},
			\mathbf{0}, \mathbf{\Omega}) +\mathcal{L}_n(\mathbf{B}_{(\mathcal{A})},
			\mathbf{0}, \mathbf{\Omega})- \mathcal{L}_n(\hat{\mathbf{B}}_{(\mathcal{A})},\mathbf{0}, \mathbf{\Omega}) \\
			&\geq \mathcal{L}_n(\mathbf{B}_{(\mathcal{A})},\mathbf{B}_{(\mathcal{A}^c)}, \mathbf{\Omega}) - 	\mathcal{L}_n(\mathbf{B}_{(\mathcal{A})},
			\mathbf{0}, \mathbf{\Omega}) >0 \ \text{in probability}.
		\end{split}
	\end{equation*}
	Therefore, we can get that $P(\hat{B}_{ij}=0, \forall i \in \mathcal{A}^c) \to 1$ and the proof of Theorem 2 is completed.
	\bibliographystyle{apalike}
	\bibliography{MH-ref}
\end{document}